\documentclass[useAMS]{mn2e}

\usepackage{epsf, epsfig, graphicx, epstopdf, float}

\title[A new model for Terzan 5]{A new model for
the multiple stellar populations within Terzan 5}
\author[M. McKenzie and   K. Bekki]
       {M. McKenzie${}^{1}$
        and K. Bekki,${}^{1}$\\
${}^1$ICRAR M468
The University of Western Australia
35 Stirling Hwy, Crawley
Western Australia 6009, Australia \\
Australia}

\begin{document}

\date{Accepted, Received ---- February 20; in original form }

%\pagerange{\pageref{firstpage}--\pageref{lastpage}} \pubyear{2018}

\maketitle

\label{firstpage}

\begin{abstract}

Recent observational studies have demonstrated that the complex stellar system Terzan 5 (Ter 5) harbours multiple populations of stars. 
Several models have attempted to interpret the large age difference of several Gyrs between the dominant populations, but none have been universally accepted. 
We propose a new scenario whereby a collision between a metal-poor Ter 5 and a giant molecular cloud (GMC) serves as a catalyst for the generation of a super-solar population of stars.
Using numerical simulations of this new ``GC-GMC'' collision scenario we demonstrate that, within a time frame of several Gyrs, our synthetic Ter 5 was capable of interacting with a metal-rich GMC in the central region of the Galaxy.
As a consequence of this, our simulated globular cluster (GC) is able to capture enough gas from the colliding GMC to form a new population of metal-rich stars.
Furthermore, the younger population created from the high-density regions of the captured gas is shown to have a stronger central mass concentration than the older metal-poor one, which is consistent with observations.
A chemical link between Ter 5 and the bulge population of the Milky Way has long been observed and these simulations finally provide evidence for their similarities. 
Our model rationalises the 5 Gyrs of quiescence observed between the two dominant populations of Ter 5 and justifies the existence of the young generation.
We discuss the advantages and disadvantages of the new scenario in the context of the observed physical properties of Ter 5.

\end{abstract}


\section{Introduction}

%%%%% TABLE1
\begin{table*}
\centering
\begin{minipage}{180mm}
\caption{The key ingredients of the new ``GC-GMC collision''
scenario. The left column lists the observed properties of Ter 5 and the right describes the physical processes responsible for the observed properties
in the new scenario. Here ``GC'' means the original massive
stellar system of Ter 5 (i.e., Ter 5's progenitor).
The detailed physical processes of the scenario are 
described and discussed in the main text.}
\begin{tabular}{ll}\hline
{The observed properties of Ter 5} &
{Physical processes } \\\hline

(1) The two distinct stellar populations  & 
Star formation from gas accreted  onto  Ter 5's progenitor system  \\
(2) Ter 5's massive progenitor system   & 
Either the nucleus of a defunct dwarf or a bulge's  massive fragment \\
(3) High  metallicity ([Fe/H]$\sim +0.2$) of the younger population  & 
Star formation from the originally metal-rich GMC  \\
(4) The $\sim  7$  Gyr age gap between the two populations & 
Rare GC-GMC bound collision (once every several Gyr)  \\
(5) Stronger central concentration in the younger population  & 
Dissipative formation of the younger population
within  the GC \\
(6) A larger number of millisecond pulsars  & 
A higher fraction of binary neutron stars in the younger population  \\
\end{tabular}
\end{minipage}
\end{table*}

Internal chemical abundance spreads in light elements (e.g., C, N, and O)
have been observed 
not only within Galactic GCs
(e.g. Norris, Freeman \& Da Costa 1983; Carretta  et. al. 2009; 
Frelijj et al. 2017), but also in 
galaxies of the Local Group  (e.g. Carretta et al. 2010; 
Larsen et al. 2001; Niederhofer et al. 2016).
However, abundance spreads in heavier elements are rarely observed in GC's,
for example, star-to-star abundance spreads in $r$-process elements
(e.g., Eu and Ba) have been observed in only a few GC's
such as M5 (e.g., Roederer 2011; Worley et al. 2013).
NGC 2808 and $\omega$ Cen
are observationally suggested to have
large internal abundance spreads in helium  ($Y$)
(e.g., Piotto et al. 2004; Norris 2004).
Only eight Galactic GCs are 
observed to have spreads in [Fe/H] 
(Marino et al. 2015; Bekki  \& Tsujimoto 2016), 
making it a substantially rarer occurrence.
The diverse nature of multiple stellar populations have been
discussed by many authors in the context of different GC formation scenarios.
(e.g., Renzini et al.  2015;  Bastian \& Lardo 2017 for a recent review).

The complex stellar system Ter 5 is located at a distance of ~5.9 kpc
from the Sun (Valenti et al. 2007) on the outskirts of the Milky Way's inner bulge. A study by Ferraro et al (2009) established the existence of two different generations; a dominant, metal-poor population with metallicities [Fe/H] ~= $-0.2$ dex and a super-solar population with metallicity [Fe/H] ~= 0.3 dex. Later, Origlia et al. (2013) discovered a very small first generation of stars with an average metallically of [Fe/H] ~= $-0.79$ dex. The most favourable interpretation of these results is that Ter 5 has two very distinct ages; 12 Gyr and 4.5 Gyr (Ferraro et al 2016). We are yet to discover another GC that shows evidence of distinct epochs of star formation linked by such a prolonged period of quiescence.

Many models have sought to explain these unique characteristics of Ter 5. It has been hypothesised that, like $\omega$ Centauri, Ter 5 is the nucleus of a defunct dwarf galaxy accreted by the Milky Way (Ferraro et al 2009; for  $\omega$  
Centauri, see Bekki \& Freeman 2003). Prager et al. (2017) argued that if Ter 5 was initially a dwarf galaxy it would have required an extremely large change in its mass to light ratio. Furthermore, Ter 5 would only be only 1\% of its original mass, placing it on the lower end of the commonly accepted mass range for dwarf galaxies after tidal stripping. (Prager et al. 2017; Stierwalt et al 2015). Additionally, when considering the [Fe/H] spread of $\omega$ Centauri (Johnson \& Pilachowski 2010) 
we see evidence of a continuous [Fe/H] distribution
with one dominant peak, whereas Ter5 has three obvious peaks (Massari et al. 2014) that are yet to be seen in other GC-like systems.

Another common theory is that Ter 5 underwent ``self enrichment" in order to induce an additional epoch of star formation (Ferraro et al 2016). It is believed that the system was assembled 12 Gyr ago, then $\sim 7$ Gyr later the super-solar component was formed from high velocity supernovae ejecta. If this was the case Ter 5's starting mass would have been at least $10^8{\rm M}_{\odot}$ (Baumgart et al. 2008) again placing it within the mass range of dwarf galaxies (Stierwalt et al 2015). Additionally, this scenario cannot easily explain why the stars which formed from SNII ejecta, are several Gyr younger than their predecessors. It is unknown as to why secondary star formation was possible so many Gyr after the original population was created.
This theory does however provide a very sound argument as to why Ter 5 harbours the largest millisecond pulsar population known to date (Ransom et al 2005). It follows that abundance of SNII would generate enough millisecond pulsars from the neutron star remnants.

In conjunction with the ``self enrichment theory'', Ter 5 could be a remnant of a pristine fragment or ``fossil remnant" of the Galactic bulge (Massari et al. 2014,  Lanzoni et al. 2013, Ferraro et al. 2016). One key observation is that the metallicity of the young bulge population of the Milky Way is strikingly similar to the younger population of Ter 5 (Lanzoni 2013 , Lanzoni et al. 2010, Feraro  et al. 2016). The younger, metal rich population is shown to have a solar scaled abundance of [$\alpha$/Fe] = +0.03 dex (Origlia et al. 2010) which is consistent with what is seen in bulge stars. Surveys such as GIBS (Gonzales et al. 2015) and the Gaia - ESO survey (Rojas - Arriagada et al. 2017) are providing evidence for multiple bulge populations, specifically a young super solar population within the bulge, which was previously a controversial topic. Our new model closely aligns with this idea.

The purpose of this paper is to discuss a new scenario of Ter 5 formation (``GC-GMC collision scenario'') in which the progenitor stellar system of Ter 5 collided with a GMC about 5 Gyr ago, creating a young stellar population in its central region.
Since Ter 5's chemical signatures are so closely related to that of the Milky Way bulge stars (e.g., Ferraro et al 2016), we believe that Ter 5's interaction with a bulge GMC triggered the rapid accretion of metal rich gas onto Ter 5, forming the younger population of stars.
Using hydrodynamical simulations, we investigate the probability of a direct GC-GMC collision which would generate secondary star formation and the physical conditions which would allow for this to occur.
Table 1 briefly summaries the key ingredients of the new scenario for Ter 5.

The main focus of this numerical study is to 
(i) describe the formation process of the metal-rich population
of stars in Ter 5 and 
(ii) reproduce the observed abundance ratios of the 2nd generation of stars. 
The present study is two-fold as follows.
We first investigate whether a collision between Ter 5 and the GMC is 
possible based on numerous orbital calculations of Ter 5 under the
gravitational potential of the Galaxy.
Then, for reasonable sets of orbital parameters derived from the previous section, investigate the dynamical and hydrodynamical evolution of Ter 5 colliding with a metal-rich GMC.
We mainly investigate 
(i) whether secondary star formation is possible within the GC and 
(ii) the spatial distribution and mass of the 2nd generation of stars.
Without further simulations we cannot show the origin of
the 3rd metal-poor population found by Origlia in (2010),
but we do propose ideas as to why this population exists. 
This model does not require Ter 5 to have unusually large initial mass
as there is no need to retain the SNeII ejecta; the large amount of gas required for the formation of the second generation of stars is accreted from the GMC. We also discuss the existence of the abnormally large millisecond pulsar population within Ter 5 in the context of the new scenario.

The structure of this paper is as follows. 
In \S 2 we describe the probability of a GC-GMC collision within the Galactic bulge ($R<3$ kpc) about 5-6 Gyr ago.
We describe the results of the new simulations of GC-GMC collisions and discuss the parameter dependence in  \S 3.
In \S 4 we outline the advantages and disadvantages of our new model and the origin of characteristic properties of Ter 5.
In particular, we discuss the origins of the third metal-poor population found by Origlia et al. (2010) and the existence of the large population of millisecond pulsars. 
Our main conclusions are presented in \S 5.

%%%%% TABLE
\begin{table*}
\centering
\begin{minipage}{165mm}
\caption{Description of parameter  for
the first set of simulations.}
\begin{tabular}{ccccccccc}\hline
{ Model ID   } &
{ $M_{\rm bulge}$ ($10^{10} {\rm M}_{\odot}$)   } &
{ $R_{\rm i}$ (kpc)   } &
{ $\theta$ (degrees)   } &
{ $f_{\rm v}$  } &
{ $M_{\rm gc}$   ($10^6 {\rm M}_{\odot}$)   } &
{ $R_{\rm gc}$ (pc)  } &
{ $N_{\rm c}$  } &
{ $N_{\rm c,b}$  } \\\hline
1  & 1 & 1 & 30 & 0.5 & 8  & 15 & 12837 & 11 \\
2  & 1 & 1 & 30 & 0.5 & 5  & 10 & 12726 & 11 \\
3  & 1 & 1 & 30 & 0.5 & 3  & 5  & 8055  & 1  \\
4  & 1 & 1 & 30 & 0.9 & 3  & 5  & 2647  & 0  \\
5  & 1 & 1 & 30 & 0.5 & 3  & 4  & 9909  & 10 \\
6  & 1 & 1 & 30 & 0.5 & 3  & 3  & 16325 & 94 \\
7  & 1 & 2 & 30 & 0.5 & 3  & 3  & 3831  & 0  \\
8  & 1 & 2 & 30 & 0.5 & 3  & 20 & 6804  & 0  \\
9  & 1 & 1 & 30 & 0.5 & 3  & 20 & 6972  & 2  \\
10 & 1 & 1 & 30 & 0.5 & 10 & 35 & 8295  & 21 \\
11 & 1 & 2 & 30 & 0.5 & 10 & 35 & 7277  & 5  \\
12 & 1 & 3 & 30 & 0.5 & 10 & 35 & 4741  & 0  \\
13 & 1 & 2 & 30 & 0.9 & 10 & 35 & 3940  & 4  \\
14 & 1 & 2 & 20 & 0.9 & 10 & 35 & 8543  & 4  \\
15 & 1 & 3 & 0  & 0.9 & 10 & 35 & 4760  & 11 \\
16 & 1 & 3 & 5  & 0.9 & 10 & 20 & 2663  & 3  \\
17 & 1 & 3 & 10 & 0.9 & 10 & 20 & 1738  & 1  \\
18 & 1 & 3 & 45 & 0.5 & 10 & 20 & 2879  & 1  \\
19 & 1 & 2 & 45 & 0.9 & 10 & 20 & 4055  & 9  \\
20 & 1 & 2 & 45 & 0.9 & 3  & 20 & 4107  & 0  \\
21 & 1 & 1 & 45 & 0.9 & 3  & 20 & 15301 & 2  \\
22 & 1 & 1 & 45 & 0.9 & 2  & 20 & 11030 & 0  \\
23 & 1 & 1 & 45 & 0.5 & 2  & 20 & 13151 & 1  \\
24 & 0 & 3 & 80 & 0.9 & 3  & 10 & 4476  & 50 \\
25 & 0 & 3 & 30 & 0.9 & 3  & 10 & 2047  & 1  \\
\end{tabular}
\end{minipage}
\end{table*}

\section{Probabilities of GC-GMC merging}

\subsection{The model}

The progenitor of Ter 5 could have been significantly more
massive that the present Ter 5, as it should have lost a significant fraction of its initial stars
through tidal interaction with the Galactic bulge
and internal dynamical evolution driven by two-body relaxation
effects (e.g., Rossi et al. 2017 for a recent work on
GC evolution in the inner part of the Galaxy).
Accordingly, it is not appropriate for us to call it ``GC'', however, we label it in this way for convenience.
In this first set of simulations, using a set of model parameters described below, we investigate whether this GC can merge with GMCs in the central region of the Galaxy.
We use our own original simulation code that can be run on GPU clusters (Bekki 2013, 2015).

\subsubsection{The Galactic potential}

The orbit of the GC is influenced
by the {\it fixed} gravitational potential of the Galaxy, which is assumed
to have three components: a dark matter halo, a disk,
and a bulge.
We assume the following  logarithmic dark matter halo potential
for the Galaxy:
\begin{equation}
{\Phi}_{\rm halo}=v_{\rm halo}^2 \ln (r^2+d^2),
\end{equation}
where
$d$ = 12 kpc, $v_{\rm halo}$ = 131.5 km ${\rm s}^{-1}$ and
$r$ is the distance from the centre of the Galaxy.
The gravitational potential of the Galactic disk is represented by
a Miyamoto-Nagai (1975) potential:
\begin{equation}
{\Phi}_{\rm disk}=-\frac{GM_{\rm disk}}{\sqrt{R^2 +{(a+\sqrt{z^2+b^2})}^2}},
\end{equation}
where $M_{\rm disk}$
is the total mass of the Galactic stellar disk,
$a$ and  $b$ correspond to 
the radial and vertical scale lengths, respectively,
and $R=\sqrt{x^2+y^2}$.
The {\it present-day} Galaxy can be modelled properly
for  $M_{\rm disk}$ = 1.0 $\times$ $10^{11}$ $M_{\odot}$
and $a$ = 6.5 kpc, $b$ = 0.26 kpc (e.g., Bekki \& Yong 2012; BY12).
The following  spherical Hernquist (1990) model is adopted for
the potential of the Galactic bulge:
\begin{equation}
{\Phi}_{\rm bulge}=-\frac{GM_{\rm bulge}}{r+c},
\end{equation}
where $M_{\rm bulge}$ is the total mass of the bulge
and $c$ is the scale length of the distribution.
BY12 adopted $M_{\rm bulge}$ =  3.4
$\times$ $10^{10}$ $M_{\odot}$
and $c$ = 0.7 kpc for the present-day bulge so that the
rotation curve profile can be properly modelled for the Galaxy.
A maximum rotation speed of 224 km ${\rm s}^{-1}$ at $R=8.5$ kpc.

A study by Massari et al. 2015 investigated the proper motion of Ter 5 and found that $(\mu_l cos(b), \mu_b) = (5.89 \pm 0.10, 0.83 \pm 0.12)$ and that Ter 5 most likely existed within a 1 kps orbit around the centre of the galaxy.
To reproduce the orbit of Ter 5 around the galaxy the following three parameters have been used:
the initial distance from the Galactic centre ($R_{\rm i}$),
the orbital inclination angle with respect to the Galactic plane
($\theta$), and the velocity factor ($f_{\rm v}$) which determines
the orbital eccentricity. The initial velocity 
of Ter 5 is given as $f_{\rm v} v_{\rm c}$,
where $v_{\rm c}$ is the circular velocity at $R=R_{\rm i}$:
$f_{\rm v}=1$ means that  the orbit is purely
circular. 
We investigate numerous models with different values of these parameter in order to find merging between GC and GMCs. 
A sample of the parameter values investigated in this study have been summarised in Table 2.

\subsubsection{GC progenitor}

Although it would be reasonable to represent the GC as a point-mass particle when investigating the probabilities of GC-GMC merging in the early Galaxy, we adopt the Plummer model for the radial density distribution of the GC.
This is so we can more accurately estimate the probability of each GC-GMC merging event within the simulation. 
The GC is represented by $10^4$ collision-less stellar particles with a size of $R_{\rm gc}$ and a mass of $M_{\rm gc}$.
We mainly investigate models with $M_{\rm gc}=3\times10^6 {\rm M}_{\odot}$ and $M_{\rm gc}=10^7 {\rm M}_{\odot}$.

\subsubsection{GMC}

The Galaxy is assumed to have numerous GMCs within the central 3 kpc.
Since it is unclear how much cold gas the Galaxy had about $[5-9]$ Gyr ago, it is reasonable for this study to assume that the total mass ($M_{\rm gmc}$) and number ($N_{\rm gmc}$) of GMCs are free parameters. 
The initial distribution of GMCs within the central 3 kpc is assumed to have an exponential profile with the scale length of 3.5 kpc. 
The apocentre of the GC orbit is always less than 3 kpc in the present study, thus we only consider the GMCs that are located within this central 3 kpc.
Each GMC is represented by a point-mass particle with a softening length of 50pc and a has circular orbit centred around the bulge. 
GMCs can gravitationally interact with one another and also be influenced by the Galactic potential and the GC.

In the present study, we mostly investigate models with $N_{\rm gmc}=3000$ for $R \le 3$ kpc.
If each GMC mass is $\approx 10^5 {\rm M}_{\odot}$, then the collective mass of the GMCs is $3 \times 10^8 {\rm M}_{\odot}$, which is only $\sim 30$\% of the present-day ${\rm H_2}$ mass of the Galaxy (e.g., Nakanishi \& Sofue 2016).
The Galaxy has $\sim 4000$ GMCs for $4 \le R \le 8$ kpc (e.g., Solomon et al. 1979) and the molecular gas shows a strong central concentration (e.g., Fig.4 of Nakanishi \& Sofue 2016).
Therefore, the above number ($N_{\rm gmc}=3000$) would be reasonable for this investigation.
It is possible that more than 5 Gyr ago the younger Galaxy could have been more gas-rich with the total 
(HI+${\rm H_2}$) gas mass being more than $10^ {10} {\rm M}_{\odot}$.
Therefore, the number of GMCs in the younger Galaxy could have been significantly larger than that of the present-day Galaxy. 
Since we do not know the gas content of the younger Galaxy, we consider that the above $N_{\rm gmc}$ is just a reference value.

Recent observations have revealed that the cumulative mass distributions of 
GMCs in the Galaxy and nearby galaxies (like M33) have a power-law form
as follows (e.g., Blitz et al. 2007):
\begin{equation}
N(m_{\rm gmc}>m_{\rm gmc}^{\prime}) \propto m^{\alpha},
\end{equation}
where $m_{\rm gmc}$ ($m_{\rm gmc}^{\prime}$) is the mass of a GMC
($10^4 {\rm M}_{\odot} \le  m_{\rm gmc}$)
and $\alpha$ ranges from $-2.49$ to $-1.71$.
The negative $\alpha$ , that is, a larger number of GMCs with lower masses, implies that low-mass GMC are more likely to collide with a GC since the probability of GC-GMC collisions is proportional to the number density of GMCs.
In order to investigate the probability of GC-GMC collisions we assume that GMCs have the same mass. 
This GMC model is over-simplified, however we believe that such a model is more than enough to discuss whether GMC-GC collisions were possible in the Galaxy.
The parameter values for GMCs are summarised in Table 2.

\subsubsection{Search for GC-GMC merging}

We examine whether there is a GMC particle near the GC at each time step in a simulation. 
If a GMC is located within $R_{\rm s}$, where $R_{\rm s}$ is the search radius from the centre of the GC, we record the GMC particle as a ``neighbour particle'' of the GC.
The total number (mass) of these neighbour particles are estimated at each time step to check if there are any GC-GMC merging candidates.
We investigate these neighbour GMC particles for $R_{\rm s}=10$, 30, and 100 pc for all models.
If there is a neighbour particle, then we investigate whether the particle is gravitationally bound to the GC using the following formula for escape velocity:
\begin{equation}
T_{\rm gmc}= 
{\Phi}_{\rm gc}(r)+0.5v_{\rm gmc}^2,
\end{equation}
where ${\Phi}_{\rm gc}(r)$ is the gravitational potential of the GC
at the neighbour's position $r$, where $r$ is the distance of the particle
from the GC centre,
where $v_{\rm gmc}$ is the velocity of the particle with respect to the
GC centre. 
If $T_{\rm gmc}$ is less than 0, the particle is considered
to be bound (labeled as a``bound collision'' or ``merging''),
meaning that the particle's velocity is less than the escape velocity relative to the GC.

To properly check the occurrence of GC-GMC merging at each time step of the simulation, $T_{\rm gmc}$, the impact parameter ($b$) and the relative velocity between the neighbour particle and the GC ($v_{\rm rel}$) is evaluated and recorded for every single neighbour particle.
These $b$ and $v_{\rm rel}$ are the key parameters for GC-GMC merging and are investigated in detail in the second set of simulations.
We run numerous simulations with a time step width of $1.4\times10^4$ yr, which is small enough for the investigation of GC-GMC merging.
A sample of the parameter values are listed in Table 2, along with the number of all collisions ($N_{\rm c}$) and bound
collisions ($N_{\rm c, b}$).

\subsection{Results}

\subsubsection{Fiducial model}

\label{Figure. 1}
\begin{figure}
\psfig{file=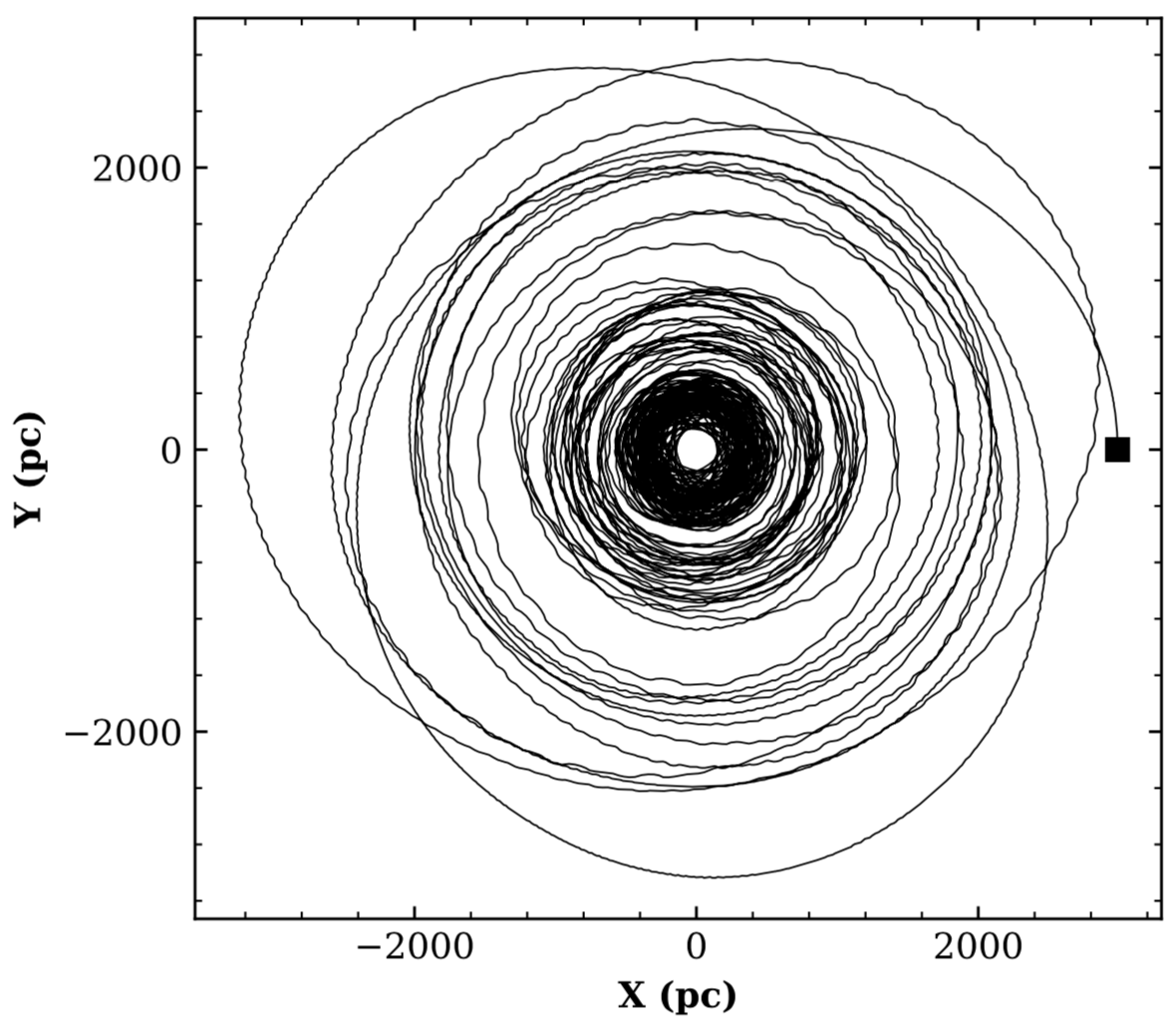 ,width=8.5cm}
\caption{
Orbital evolution of the GC in model 16 projected onto the X-Y plane. The starting position is denoted by a black square, 3 kpc away from the centre. 
As time progresses, the orbit of the GC decays into a more stable orbit 1 kpc 
from the centre owing to dynamical friction of the GC against low-mass
GMCs.
}
\end{figure}

\begin{figure}
	\psfig{file=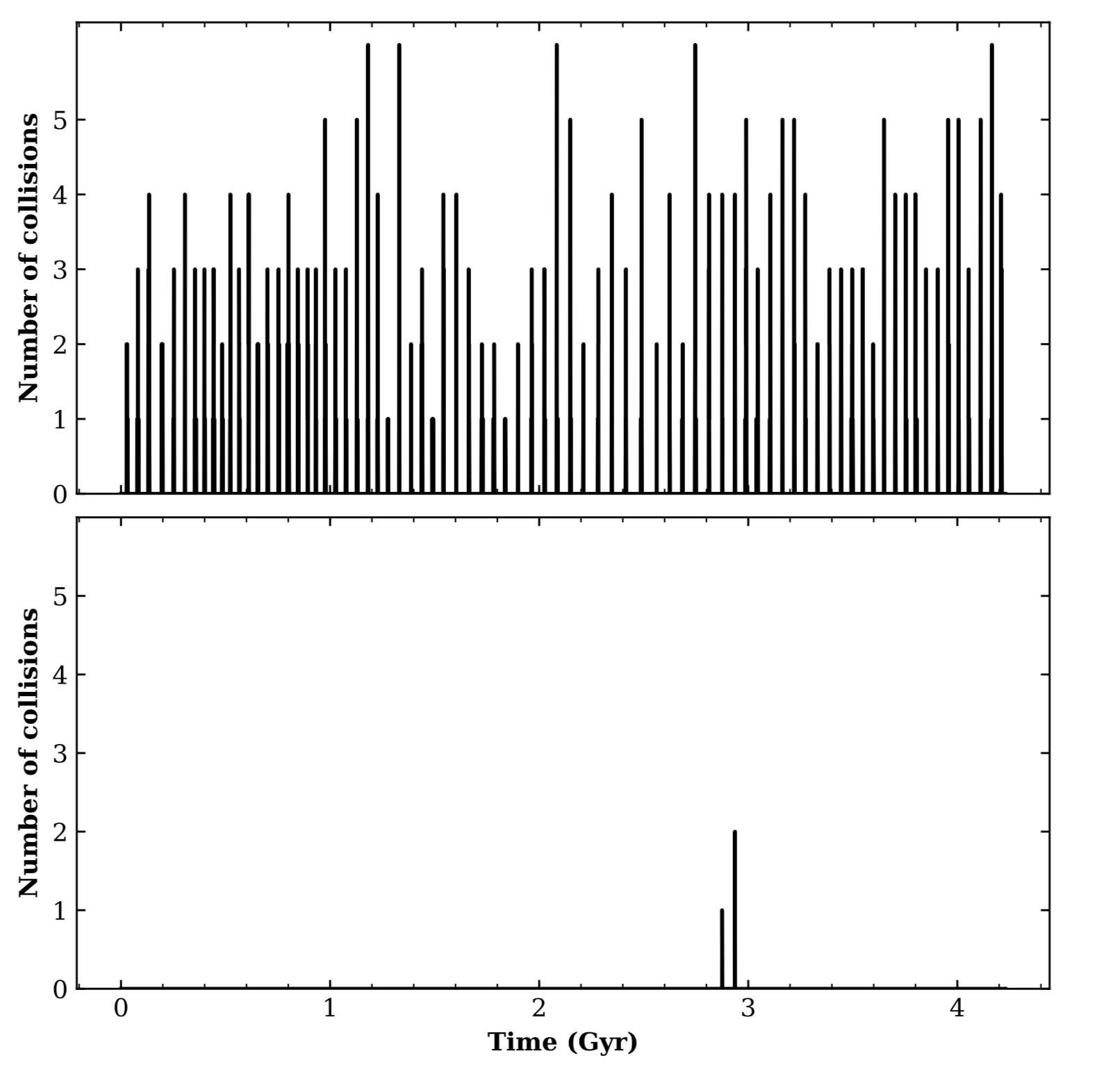,width=8.5cm}
	\caption{
	Time evolution of the number of all collisions (upper) and for the gravitationally bound collisions (lower) for model 16. The three bound collisions makes this a good model for further investigation.
}
\label{Figure.2}
\end{figure}

\label{Figure. 3}
\begin{figure*}
\psfig{file=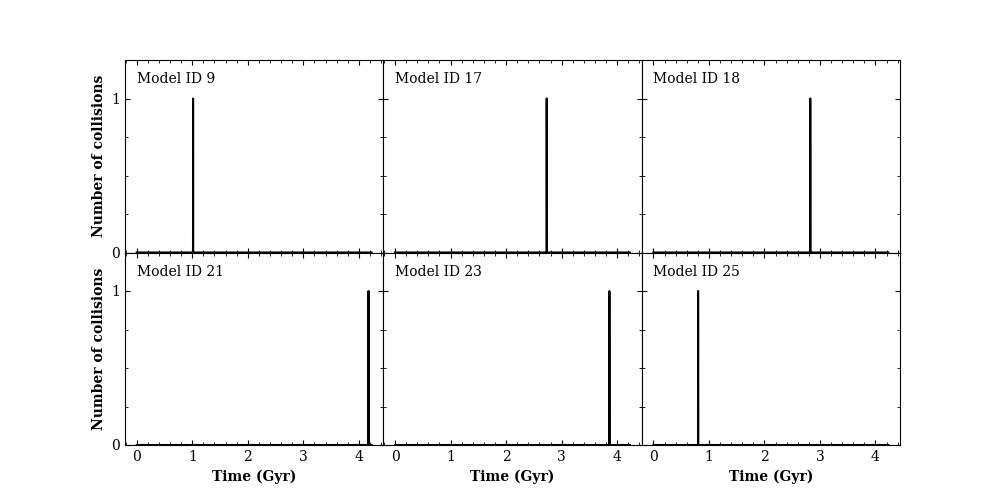 ,width=18cm}
\caption{
Time evolution of bound GC-GMC collisions in the six representative models
among all models summarised in Table 2. Note that for points where there are two collisions they occurred too close together to visually identify each one.
}
\end{figure*}

The orbits of the GC for each simulation were highly variable and primarily depended on the 
three parameters  $R_{\rm i}$, $\theta$ and $f_{\rm v}$. 
Accordingly, the GC-GMC interaction history depends on these three parameters.
Here we first describe the results of the fiducial model with model ID 16 (M16),
though numerous orbital models of the GC (Ter 5's progenitor) have been
investigated.
Our simulations show that it is possible for the GC to collide with GMCs in the inner
region of the Galaxy ($R<3$ kpc), however, almost all collisions are not the ``bound collision'' in which
GMCs can be captured by the GC so that a large amount of gas can be accreted onto the GC. 
Only one bound collision with the GMC is necessary to spawn the new metal-rich second-generation population in Ter 5, thus we consider that the models which best represent the orbital evolution of Ter 5 are those which minimise the number of gravitationally bound collisions.
If multiple GC-GMC collisions can occur at more than two totally different epochs, then the GC can have multiple episodes of star formation.
This would result in more than two populations with significant age differences, which is inconsistent with observation.
We have chosen M16 as the fiducial model as it demonstrates that a bound GC-GMC collision is possible within a short time frame.
As summarised by Table 2, there are a range of parameters that generate small numbers of gravitationally bound collisions.

Fig. 1 shows that the orbital apocentre and pericentre of the GC in our fiducial model become smaller as the GC moves around the Galaxy.
This is because the GC can lose its initial orbital angular momentum and kinetic energy through dynamical friction against GMC particles in the inner Galaxy. When considering the probability density plot made by Massari et al 2015, our orbital evolution agrees with the with a high probability of the GC residing within a 1kpc radius of the inner galaxy.
Fig. 2 demonstrates that numerous collisions ($N_{\rm c}\sim 2700$) between the GC and GMCs are possible during the orbital decay of the GC in this model. We see that the frequency of collisions is not related to the number of bound collisions. This observation is consistent with all other models that we ran. Almost all of the collisions have a large impact parameter ($b$) and a rather high relative velocity. Therefore, such unbound collisions can not trigger gas accretion from GMCs onto the GC.
Fig. 2 also demonstrates that the GC can experience three GC-GMC collisions within
a very short time scale of less than 0.1 Gyr. These three are promising GC-GMC 
collisions in which a large amount of GMC gas can be captured gravitationally
by the GC.

In the fiducial model, the original GC has a mass of $10^7 {\rm M}_{\odot}$, which is much larger than Ter 5's present day mass of  $2\times10^6{\rm M}_{\odot}$ (Lanzoni et al. 2010).
Webb \& Leigh (2015) demonstrated that originally GC€™s were on average 4.5 times larger than their present day mass which places $10^7{\rm M}_{\odot}$ well within the realm of possibility. 
For our simulations we have explored a number different parameters, but we here focus mainly on masses of $10^7{\rm M}_{\odot}$ and $3\times10^6{\rm M}_{\odot}$.  
We included models with a GC mass of $3\times10^6 {\rm M}_{\odot}$ to allow for less efficient tidal stripping of GC stars.
The lowest possible mass that resulted in collisions was found to be $2\times10^6 {\rm M}_{\odot}$, whereas the highest mass with less than three collisions was $2\times10^7 {\rm M}_{\odot}$. 
These bounds set a very large range for the possible initial mass of Ter 5, thus adding to the validity of our model.

The tidal radius of Ter 5 is observed to be quite small for a GC of this size, only 267" or 7.9 pc (Massari et al. 2014). The radii listed in Table 1 should not be confused with this tidal radii, but we found them to not make any drastic changes into the number of gravitationally bound collisions. A greater investigation into the implications of the radii is found in section 3.

For models where there are minimal collisions ($<3$) we analyse the the collision parameters, namely the relative velocities between the GC and GMC $V_{rel}$, and the impact parameter $b$. The results from these collisions are summarised in Table 3. These values are very important, as they form the basis of the second part of the simulations.  
Fig. 4 demonstrates the exponential nature of $b$ and $V_{rel}$; higher $b$ generally have lower $V_{rel}$ and visa versa. 
Generally only collisions with $V_{rel}$ $< 30$ km s$^{-1}$ and $b$ $< 50$ pc will be suitable for the next set of simulations.

\subsubsection{Parameter dependence}
We ran several different models to evaluate the impact of each parameter on the outcome of the simulations. Twenty-five of these models have been included in Table 1 and are representative of the typical outcomes of our simulations. Firstly we experimented with two different 
Galaxy models; one with a bulge and one without. 
We investigated both models as it is unclear whether a classic, massive, spherical bulge existed in the Galaxy over 5 Gyrs ago.
Furthermore, the central bar in the inner Galactic disk could correspond to the ``bulge'' component of the Galaxy.

A profound observation of our simulations was that the presence of a bulge suppressed the number of gravitationally bound collisions. 
When the bulge was removed from the simulation we found many more bound collisions compared to the bulge-less simulations. M24 and M25 are examples of the bulge-less model; when the same parameters were used with a bulge no bound collisions were found. For M25, the collision parameters $V_{\rm x}$ and $b$ were analysed and were found to be comparable to all other models run (see Table 3). We decided not to pursue these models due to lack of evidence for a bulge-less model of our Milky Way.

A larger starting position of the GC, $R_{\rm i}$,  minimises the time required for the GC's orbit to decay into the centre where the majority of the GMC's exist, thus resulting in a higher number of gravitationally bound particles. The angle of inclination above the Milky Way disc makes a minor difference to the number of bound particles; at 45$^{\circ}$ we receive the least number of collisions and the most at 0$^{\circ}$ and 90$^{\circ}$. M13 and M14 demonstrate how a small change in the inclination made no difference to the number of bound collisions. It was noticed that the velocity factor $f_{\rm v}$ was inversely proportional to the GC mass.

We experimented with a velocity factor of $f_{\rm v}$ = 0.5 and $f_{\rm v}$ = 0.9 and found a lower $f_{\rm v}$ can increase the chances of a collision between the GC and a GMC.
When high mass GC's such as $10^7{\rm M}_{\odot}$ were used we found that there were often too many gravitationally bound particles and $f_{\rm v}$ = 0.9 was beneficial in reducing that number. 
For lower mass models like $3\times10^6{\rm M}_{\odot}$, which struggled to have any bound collisions with GMCs in the bulge, the lower $f_{\rm v}$ = 0.5 increased the chances of a collision.
It was noted that the number of collisions was proportional to the number of bound collisions, but otherwise there was no discernible relationship between the two values.

Table 3 demonstrates that a number of suitable models with $<3$ gravitationally bound collisions has been found, but there is no discernible correlation between the initial conditions of the simulation and the impact parameters.
Furthermore, as shown in Fig. 3, the epoch of bound GC-GMC collision are somewhat random. 
Fig. 4 clearly demonstrates that only a fraction of these bound GC-GMC collisions have small enough impact parameters ($b<50$ pc, less than size of high-mass GMCs) and relative velocities ($v_{\rm rel}<30$ km s$^{-1}$) to be considered for further simulations.
The four outliers to the right of Fig. 4 come from small, compact masses but none of the parameters appear to effect these results as they appear to be completely random.

The results from our numerical simulations demonstrate that although bound GC-GMC collisions are rare they are indeed possible within a timescale of less than several Gyrs.
Our GC-GMC collision scenario can explain the presence of the younger metal-rich population of Ter 5 in the context of a rare GC-GMC bound collision, and considering that Ter 5 is only the Galactic GC we know of that has two distinct stellar populations with a large age difference ($\sim 7$  Gyr), our explanation is not unreasonable.
However, the scenario can not explain why the younger population is $\sim 5$ Gyr old given that there is no preferred epoch of bound GC-GMC collisions. 
It is possible that the Galaxy had a significantly higher molecular gas density about 5 Gyr ago for some physical reasons; for example, a stellar bar could have been formed 5 Gyr ago to induce gas infall to the bulge region of the Galaxy. 
Such a higher molecular gas density could have enhanced GC-GMC bound collisions owing to a larger number of GMCs in the Galaxy's center.
It is beyond the scope of this paper whether the Galaxy could have a significantly larger amount of molecular gas in its bulge region.

\begin{table}
\centering
\begin{minipage}{8.5cm}
\caption{Description of impact parameters ($b$) 
 and initial relative velocities ($V_{\rm x}$)
from collisions where there was a low number of gravitationally bound particles.}
\begin{tabular}{ccc}\hline

{ Model ID   } &  { $V_{\rm rel}$ (km/s)   } &  { $b$ (pc)   }  \\\hline
3  & 9.70  & 93.9 \\
9  & 13.4 & 53.3 \\
   & 12.8 & 74.3 \\
16 & 20.2 & 40.9 \\
   & 28.9 & 39.9 \\
   & 24.5 & 62.6 \\
17 & 27.7 & 70.6 \\
18 & 30.2 & 31.0 \\
21 & 14.2 & 62.7 \\
   & 10.4 & 99.0 \\
23 & 5.00    & 36.5 \\
25 & 7.50  & 57.8
\end{tabular}
\end{minipage}
\end{table}

\begin{figure}
\psfig{file=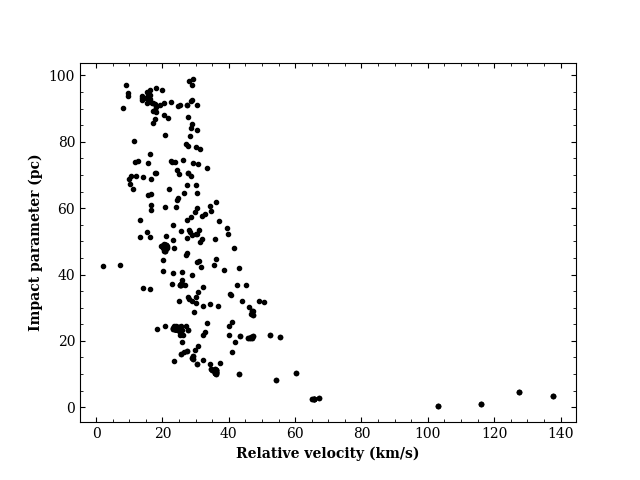,width=8.5cm}
\caption{
Impact parameters of GC-GMC collisions
as a function of the relative velocities between the GC and GMCs
for collisions listed in Table 1.
}
\label{Figure.4}
\end{figure}

\section{Numerical simulations of GC-GMC merging}

\subsection{The model}
The first set of simulations demonstrated that GC-GMC collisions with small impact parameters and relative velocities are possible for a range of parameters.
Next we consider whether the Ter 5 progenitor can merge with a fractal GMC with a mass of $m_{\rm gmc}$ and a size of $R_{\rm gmc}$ in the central region of the Galaxy.
We adopt the fractal GMC model used in Bekki (2017; B17) for the present study, the details of which have already been outlined in B17. 
Thus we only briefly describe the model here.

\subsubsection{Fractal GMC}

We consider that the GMC has a power-law radial density profile
($\rho_{\rm gmc}(r)$) as follows:
\begin{equation}
\rho_{\rm gmc} (r)=\frac{\rho_{gmc,0}}{ (r+c_{\rm gmc})^{\beta} },
\end{equation}
where $r$,  $\rho_{\rm gmc, 0}$,  and $c_{\rm gmc}$,
$\beta$  are the distances from the GMC's
center, a constant that is determined by $M_{\rm gmc}$ and $R_{\rm gmc}$,
the core radius of the GMC, and the power-law slope.
Guided by the observed range of $\beta = 1-2$ (e.g., Ashman \& Zepf 2001),
we adopt $\beta=1$ as a reasonable value. This 
means that the total mass of a GMC is roughly proportional to
$R^{3-\beta}$, for which $\beta=1$ is consistent with the observed
mass-size relation ($m_{\rm gmc} \propto R_{\rm gmc}^2$).

A GMC is assumed to have a fractal gaseous distribution characterised by
a fractal dimension $D_3$. The details of a way to set up the initial
condition of a fractal structure for a given $\beta$ are given in
B17. 
We adopt $D_3=2$ as a reasonable value
for  $\beta=1$ in all models of the present study.
The initial viral ratio ($t_{\rm vir}$) of the GMC is set to be 1 and initial rigid rotation is not considered.
The GMC is represented by $N=1048911$ particles in the fiducial high-resolution model.
In order to find the best model which mimics the observed properties of Ter 5 we first ran numerous low-resolution models with $N<250,000$. For the most realistic models which were found we then ran high-resolution simulations to better explain the observations made in the low-resolution simulations.

\subsubsection{Star formation and SN feedback}

Collision-less new stellar particles (`new stars') are created from a gas particle if the following two physical conditions can be met.
Firstly, if the local density ($\rho_{\rm g}$) exceeds a threshold density
(${\rho}_{\rm th}$) for
star formation:
\begin{equation}
{\rho}_{\rm g} > {\rho}_{\rm th}.
\end{equation}
star formation is observed to proceed in the dense cores of GMCs and accordingly,
$\rho_{\rm th}$ is set to be $10^5$  H atoms cm$^{-3}$,
which is consistent with the observed values (e.g., Bergin \& Tafalla 2007).
Secondly, the local velocity field around a gas particle
is consistent with that for gravitationally collapsing,
which is formulated as follows:
\begin{equation}
div  {\bf v}<0 .
\end{equation}
One single gas particle is converted into just one new star in the present
study and can not dramatically increase during a simulation.
Although feedback effects from supernovae and AGB stars are included in the present simulations, the AGB wind is not necessary in a simulation where only 14 Myrs of evolution is investigated.

\subsubsection{The GC}

The GC with a mass of $M_{\rm gc}$ and a size of $R_{\rm gc}$ is represented by $N=10^5$ collision-less stellar particles and is assumed to have a Plummer density profile with a scale length of $0.2R_{\rm gc}$. 
The initial position of the GC with respect to the merging GMC (${\bf x}$) is as follows:
\begin{equation}
{\bf \rm x}= ({\rm 130 pc}, b, 0),
\end{equation}
where $b$ is the impact parameter.
The initial 3D velocity of the GC ($v$) is as follows:
\begin{equation}
{\bf \rm v}= (-v_{\rm rel}, 0, 0),
\end{equation}
where $v_{\rm rel}$ is the initial relative velocity between the GMC and the GC.
We consider the $b$ and $v_{\rm rel}$ derived in the first set of simulations when choosing the values of the two parameters but we also investigate models where $b$ and $v_{\rm rel}$ do not originate from our results.

\subsection{Results}

\subsubsection{Fiducial model}

\begin{table*}
\centering
\begin{minipage}{180mm}
\caption{Results for our two fiducial models during the second phase of simulations.}
\begin{tabular}{cccccccccc}\hline
Model ID & 
{ $M_{\rm gc}$   ($10^6 {\rm M}_{\odot}$)   } & 
{ $R_{\rm gc}$ (pc)  } &
{ $b$ (pc)  } &
{ $V_{\rm x}$ (km/s)   } & 
{ $M_{\rm gmc}$   ($10^6 {\rm M}_{\odot}$)   }  &
{ $R_{\rm gmc}$ (kpc)  }  & 
{ $M_{\rm old}$   ($10^6 {\rm M}_{\odot}$)   }  & 
{ $M_{\rm new}$   ($10^6 {\rm M}_{\odot}$)   }  &  \\\hline
1  & 10 & 20 & 40 & 20 & 1   & 0.1  & 2.52 & 0.14 \\
2  & 3  & 20 & 20 & 10 & 1   & 0.06 & 1.23 & 0.22    \\

\end{tabular}
\end{minipage}
\end{table*}

\begin{figure*}
\psfig{file=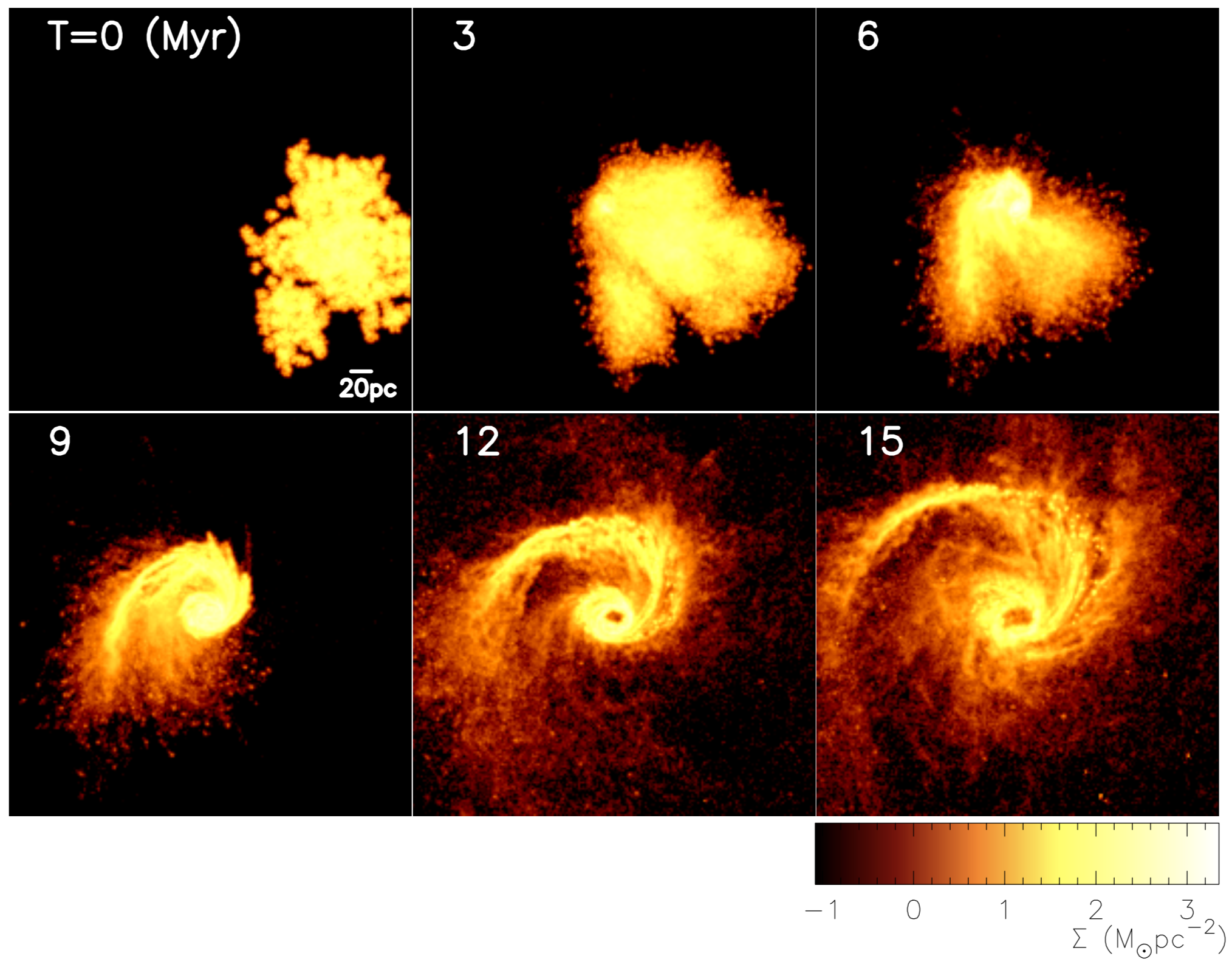 ,width=16cm}
\caption{
Time evolution of the projected mass density of 
the gas ($\Sigma$ ${\rm M}_{\odot} p^{-2})$) for the fiducial model M1. 
The centre of each frame is coincident with the centre of the GC,
although the GC is not shown in this figure.
The time T that has elapsed since the simulation started is shown 
in the upper left corner of each panel.
}
\label{Figure. 5}
\end{figure*}
\begin{figure*}
\psfig{file=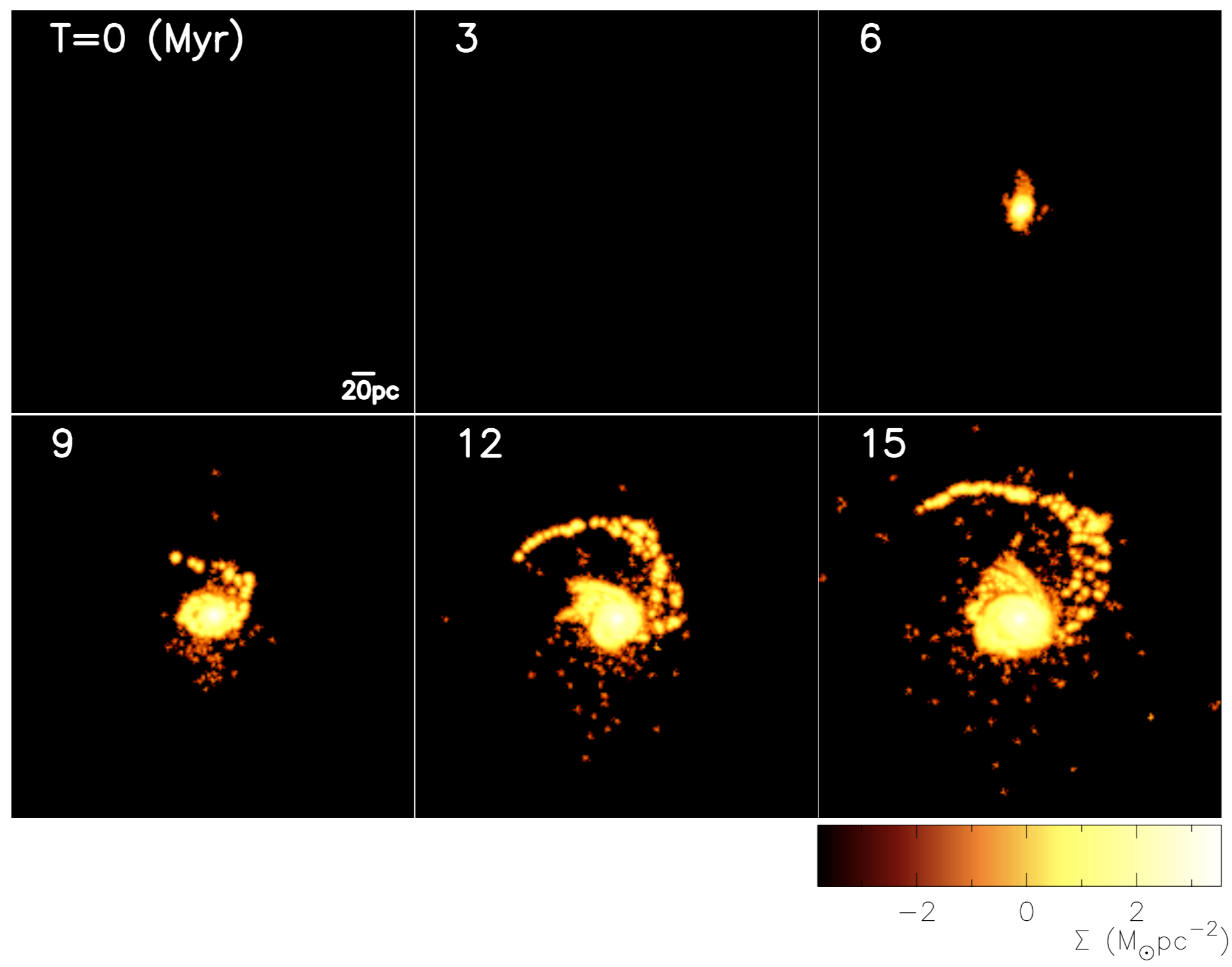 ,width=16cm}
\caption{
The same as Fig. 5 but for the new stars.
The centre of each frame again coincident with the centre of the GC.
It is clear that numerous low-mass clusters are formed in the tidal
tail that was developed during the GC-GMC collision.
}
\label{Figure. 6}
\end{figure*}

We describe the results of the fiducial model (M1, Table 4) in this second set of simulations since this model clearly demonstrates that new stars (second generation of stars, corresponding to the metal-rich population of Ter 5) can be created from a collision of the GC with the GMC. 
In order to find the best possible model, we have run more than 100 low-resolution simulations with different combinations of parameters. 
Since many of the models do not show enough consistency with physical observations, we briefly describe their results in the Appendix instead. 
The Appendix also discusses a multitude of parameter combinations that can produce centrally concentrated collections of new stars, but are not always consistent with the fiducial model.
This fiducial model adopts the orbits of the GC and the GMC from the previous set of simulations (M16, Table 2) as it has been demonstrated that  a bound GC-GMC collision (i.e., GC-GMC merging) is possible within a time-scale of less than 4 Gyrs.
One of the three gravitationally bound collisions (as seen in Fig. 2 \& Table 2) is chosen to be investigated here:
$V_{\rm x}$ = 20km/s (previously $V_{\rm rel}$)and $b$ = 40 pc. The other two collisions have been excluded as they have collision parameters not conducive to producing a large number of new stars.

Figs. 5 and 6 describe the time evolution of the mass density for the gas and new stars respectively, projected onto the $x$-$y$ orbital plane of the colliding GC-GMC pair.
The evolution of old stars in the GC has been excluded in these figures as it remains relatively constant throughout the simulations. 
As time progresses, the fractal cloud of the GMC in Fig. 5 is severely disturbed by the strong gravitational field of the GC.
A significant fraction of the GMC is tidally captured by the GC during the collision, and consequently the dense gaseous region is formed within the inner region of the GC.
After 6 Myrs, the densest region of the colliding GMC starts being converted from gas particles into new stars (Fig. 6, T=6).
A significant fraction ($\sim 30$\%) of the original gas can be converted into new stars within less than 10 Myrs.
About 80\%  of the new stars are located in the central region of the GC and these new stars correspond to the metal-rich population of Ter 5.

Owing to the strong tidal force of the GC, the GMC forms a single tidal arm which rotates around the central region. 
As seen in Fig. 6, new stars start forming within this arm and could either be tidally stripped or could fall back into the cluster as time progressed. 
The new stars in the tidal arm appear to form low-mass clusters, most of which may well be disrupted by the tidal field of the Galactic bulge in the future.
Our numerical simulations are the first to demonstrate the formation of metal-rich low-mass clusters during a GC-GMC collision. These clusters serve as a fossil record of past GC-GMC collision during the creation of Ter 5. 
If observational evidence of these clusters within the Galactic bulge was found it would greatly strengthen the validity of our model.
The final snapshot of the GMC shows a central star-deficient region where the gas has been either fully transformed into new stars or blown away by SN wind.

Fig. 7 shows that star formation begins roughly 3.5 Myr after the commencement of the GC-GMC interaction. 
This is also observed in Fig. 6 between T=3 and T=6.
Star formation in the colliding GMC soon becomes ``bursty'' with a maximum rate of $~0.7 {\rm M}_{\odot} / {\rm yr}$, because a large fraction of the gas is compressed and then captured by the central region of the GC to form dense gaseous region. The initial spike occurring at 4 Myr is largely due to the dynamical impact of the GC-GMC collision.
This model shows an extended period of up to 12 Myr of star formation, which can be attributed to the compression of the arm formed by the GMC.
Although the long-term evolution ($\sim 1$ Gyr)  of this GC-GMC collision is not investigated in the present study, it is highly likely that the residual star formation can be truncated by later physical processes in the Galactic bulge, such as tidal or ram pressure stripping of the gas from the GC by the bulge and energetic SN wind.

\begin{figure}
\psfig{file=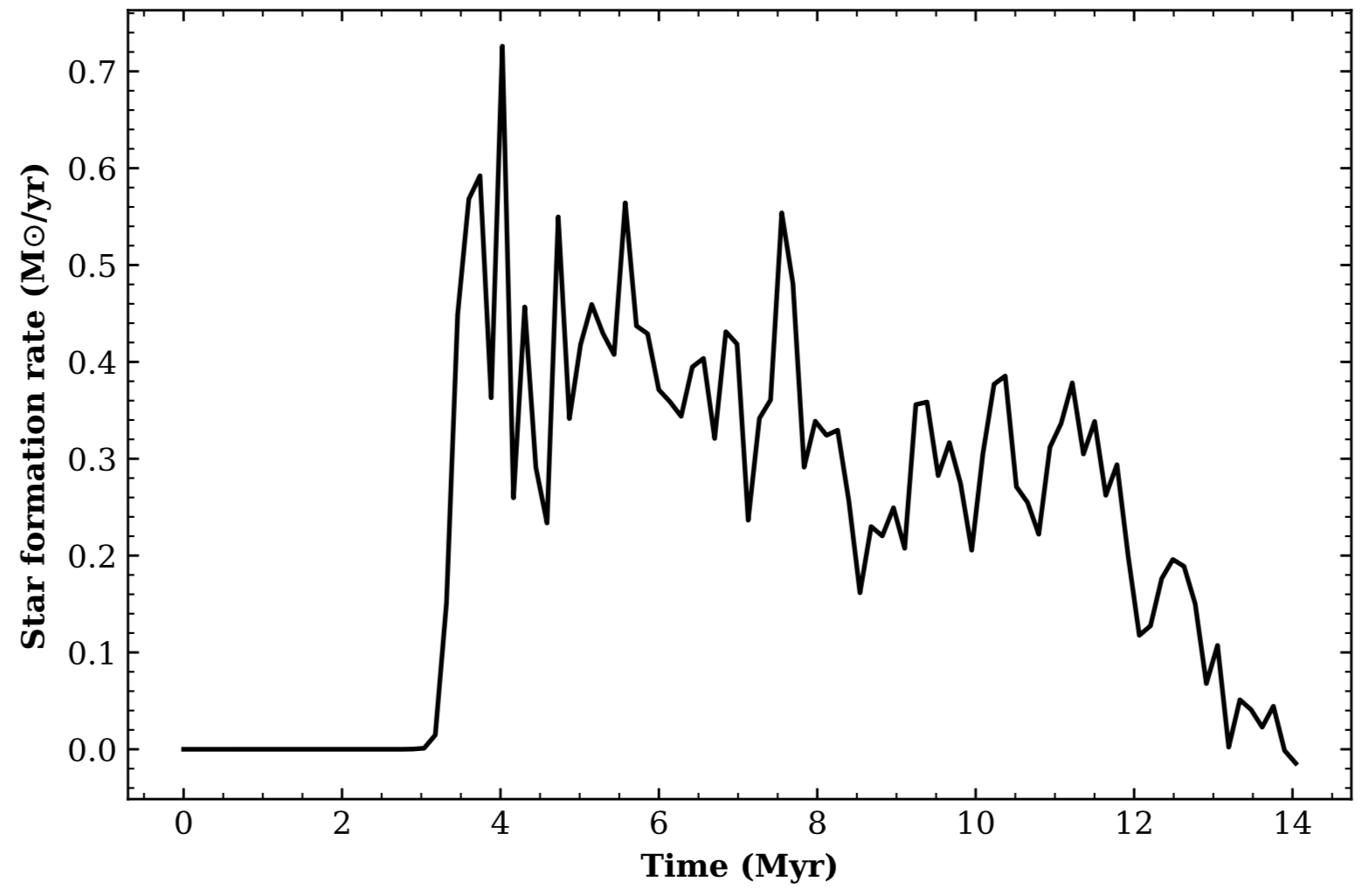,width=8.5cm}
\caption{
Star formation rate for the fiducial model M1. 
}
\label{Figure.7}
\end{figure}
\begin{figure*}
\psfig{file=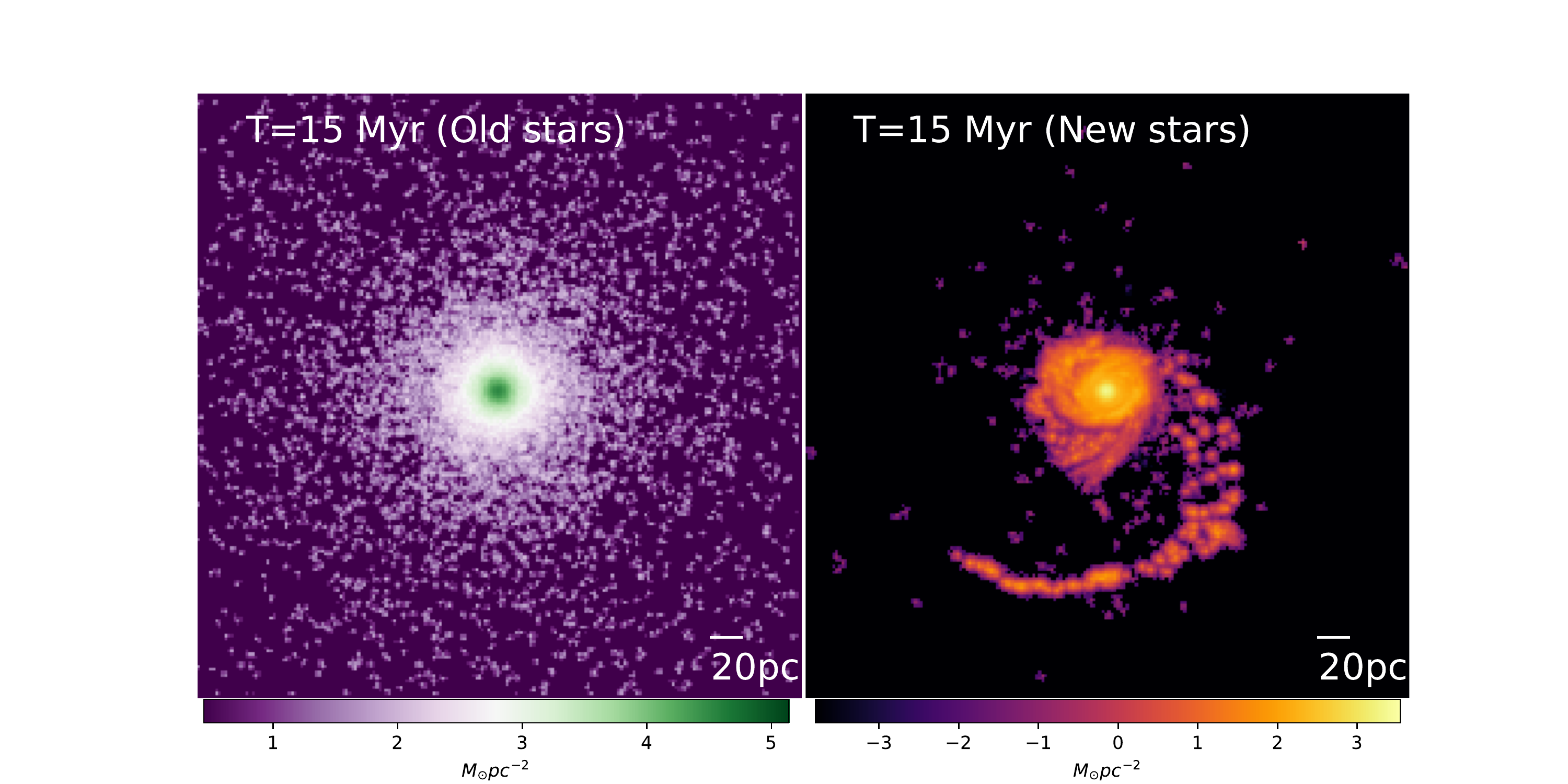,width=18cm}
\caption{
Final surface mass densities of stars in  M1 with old stars on the left and new stars on the right.
The two 2D density maps are rotated by 180 degrees and flipped over
so that numerous star cluster in the long tidal tail from the destroyed
GMC can be more clearly seen (i.e.,
the long tail of stars in the 2D distribution of new stars can not
be on top of the characters for the time $T$).
}
\label{Figure.8}
\end{figure*}

Fig. 8 shows that old stars from the original GC are spherical in nature with a maximum density of 
$~10^{4.5}  {\rm M}_{\odot} {\rm pc}^{-2}$ in the projected mass distribution.
The new stars have a slightly lower maximum density of $~10^3  {\rm M}_{\odot} {\rm pc}^{-2}$.
They also show a very clumpy distribution in the tidal arm, and the stars in the arm appear to be a collection of low-mass star clusters.
This result implies that metal-rich star clusters can be formed from GC-GMC collisions.
After the collision, the GC could undergo dynamical mass segregation and achieve a density of around $(1-4)\times10^6 {\rm M}_{\odot} {\rm pc}^{-2}$ as listed in Lanzoni et al. 2010. 
Since the long-term dynamical evolution driven by two-body relaxation, which would result in the mass segregation, is not investigated in the present simulations, it remains unclear whether such mass segregation can be confirmed.
Approximately 34\% of the particles that make up the cloud have been converted into new stars in this fiducial model.

Fig. 9 shows a plot of the cumulative distribution functions of old and new stars in this fiducial model. 
Ferraro et al. 2009 first modelled the distribution of stars in Ter 5 in this way to illustrate the central concentration of new stars. 
In their plot they give the maximum radius as 40 arc seconds. 
Using the conversion of 100" to 2.86 pc given by Massari et al. 2014, to recreate this plot we would need to use a radius of $~$2 pc. 
There simply aren't enough old stars present within this small radius to achieve a smooth plot, which seems at odds with our density estimates. Within this smaller radius there was a total mass of $1.23\times10^5{\rm M}_{\odot}$ for the old stars and $0.15\times10^4{\rm M}_{\odot}$. 
This mass is obviously too low to be representative of the present day Ter 5. 
We have chosen the tidal radius of 8 pc to be the bounds of our distribution as per the observations given by Massari et al. 2014. Within this radius there is obvious evidence that the new stars are more centrally concentrated. 
We found that there was $2.52\times10^6{\rm M}_{\odot}$ old stars and $1.38\times10^5{\rm M}_{\odot}$ new stars within the tidal radius of our GC. 
Given the different scale in the x-axis, our plot can not be directly compared to the one made by Ferraro et al. 2009.

\begin{figure}
\psfig{file=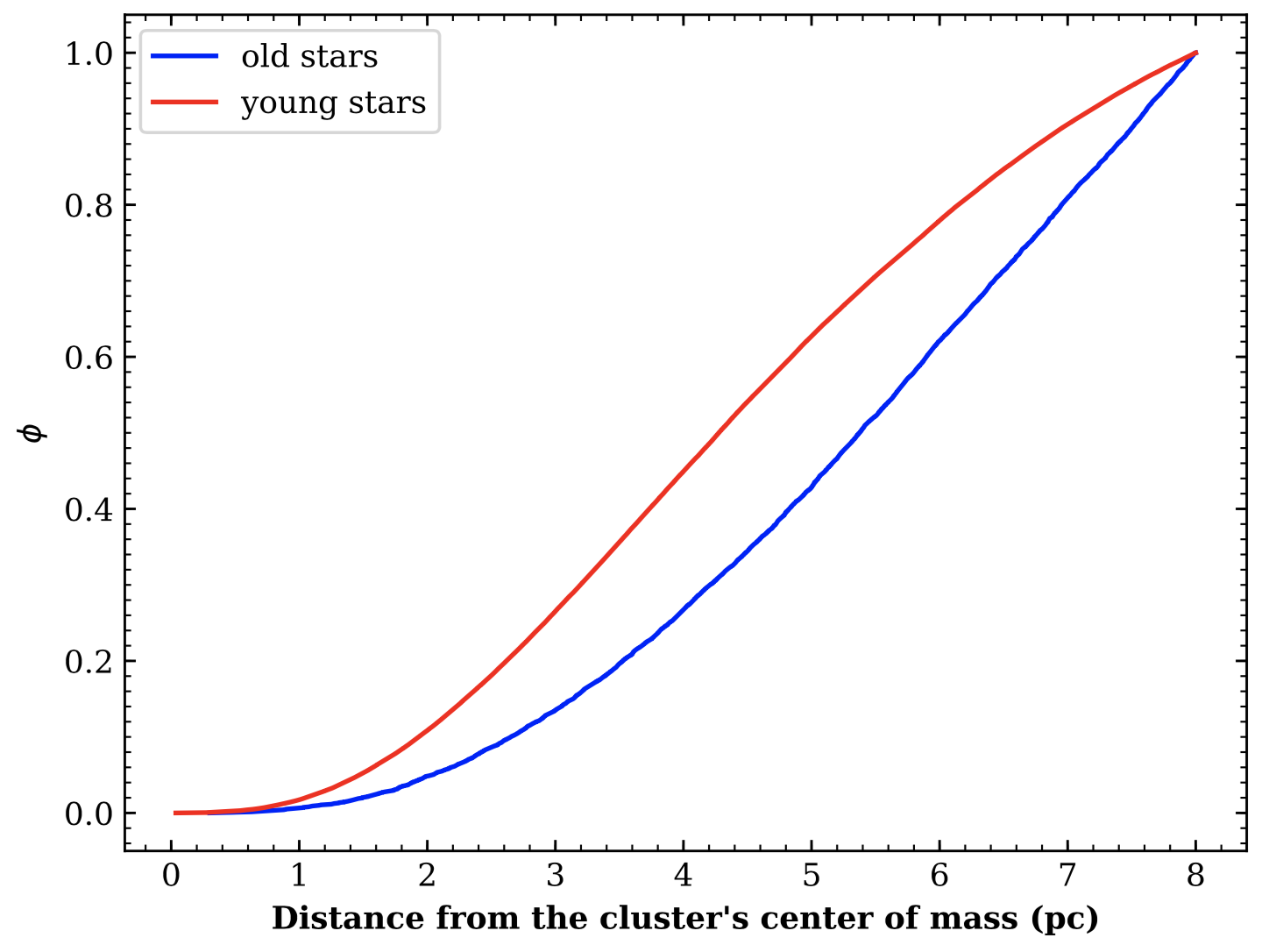 ,width=8.5cm}
\caption{
The cumulative distribution function within a radius of 8 pc for M1. New stars are featured in red and old stars in blue.
}
\label{Figure. 9}
\end{figure}
\begin{figure*}
\psfig{file=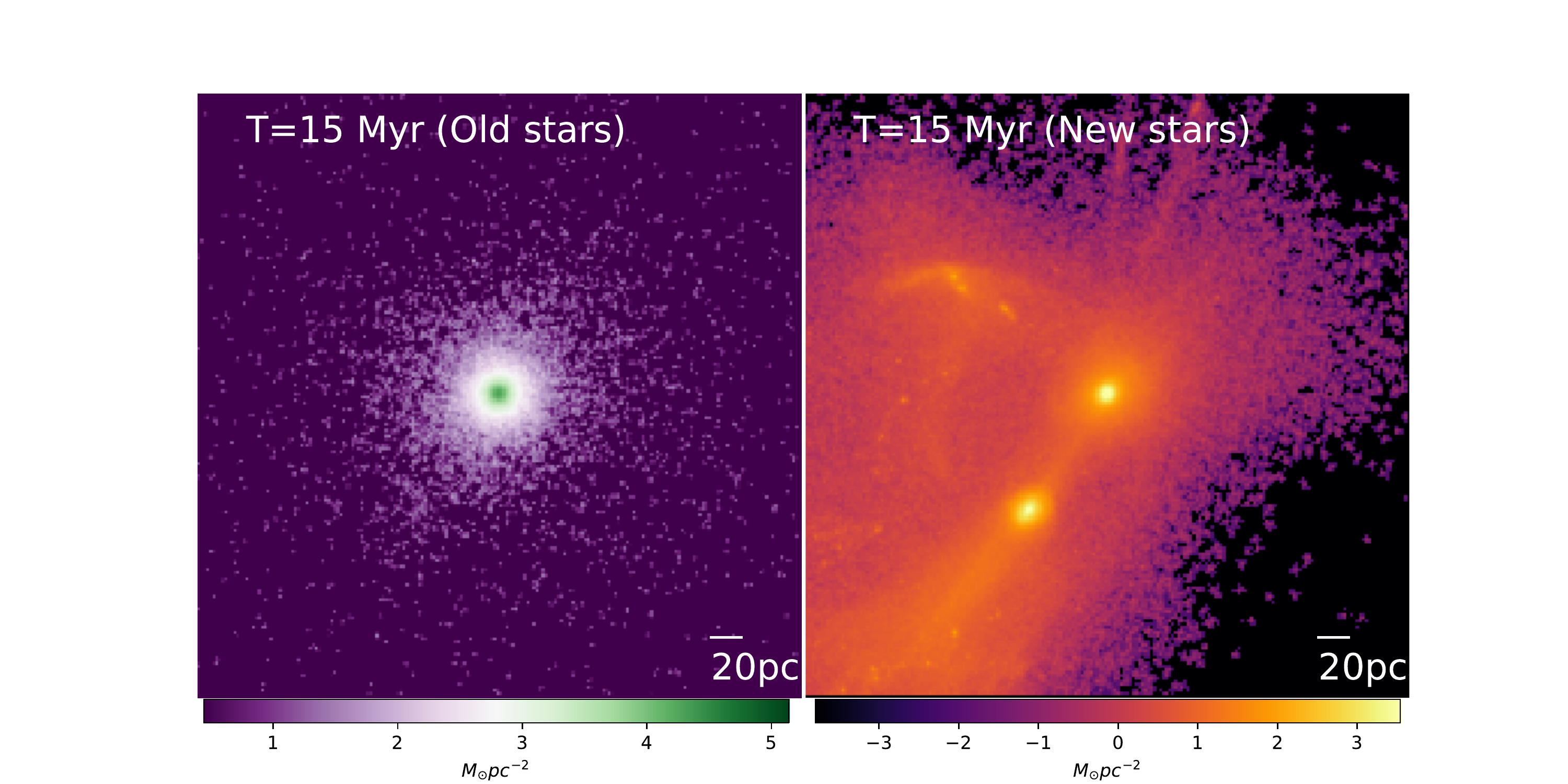 ,width=18cm}
\caption{
The same as Fig. 8 but for M2.
}
\label{Figure. 10}
\end{figure*}

The radial profiles of the line of sight velocity were analysed for each individual population of the simulated fiducial GC. We expect that there is some form of rotation within the new stars of M1 as the spiral arm in Fig 6 appears to have some net clock-wise rotation. Analysis of the line of sight rotation gave rise to the conclusion that both populations are rotating about the clusters combined centre of mass.

Massari 2014 estimated the velocity dispersion of Terzan 5 to be $\approx$15 km/s. The velocity dispersion of simulated clusters was measured by selecting stars only within the half-mass radius of the GC. This was necessary as the random motions of stars which had left the gravitational influence of the GC would lead to a higher than necessary velocity dispersion. M1's velocity dispersion for the young population was 19.04 Km s$^{-1}$, the old population was 19.36 Km s$^{-1}$ giving a total dispersion of 19.07 Km s$^{-1}$ for the whole cluster. Given the large initial size of the simulated GC, the larger than average velocity dispersion could be considered reasonable. A velocity dispersion profile of M1 shows that the new stars in the central 10pc of the cluster have a much lower velocity dispersion a much stronger net rotation than at the tidal radius. This is opposite to the older population which has the highest velocity dispersion at its centre and decreases with increasing radius.

As an alternative to our fiducial model, we have included a simulation of a lower mass GC with a higher mass GMC referred to as M2.
The model M2 shows a stronger concentration of new stars than seen in M1.
The most notable difference between this model and the fiducial model
is a large change in mass from $10^7{\rm M}_{\odot}$ to $3\times10^6{\rm M}_{\odot}$. 
Additionally, the collision parameters of  $V_{\rm x}$ = 20km/s and $b$ = 10 pc are hypothetical and did not come from the first set of
simulations described in Section 2. 
However, we do think that the adopted parameters for M2 were indeed possible for the real evolution of Ter 5.
The density of the GMC has also been increased owing to the decrease in radius from 100 pc to 60 pc. 
An increase in GMC density has been observed to increase the efficiency of the conversion of gas to stars, thus for this model where the initial GC mass is lower, the high density GMC has been used.

The time evolution of M2 is seen to proceed in a very similar way to M1.
Fig. 10 presents the final mass distributions of old and new stars of the GC in this model. 
An interesting characteristic of high density GMC models is that secondary clusters of new stars often formed around the GC's nucleus.
This phenomenon can be observed in the lower left portion of the right panel in Fig. 10. 
Owing to the high abundance of gas particles and lower gravitational potential well of the GC, not only has a central cluster formed within the old stars of the GC, an additional cluster of comparable density has also formed outside the GC's tidal radius. 
As discussed in Appendix A, multiple new star clusters are more pronounced in 
even lower mass models (e.g., M56 with a mass of $2\times10^6{\rm M}_{\odot}$ generated an additional three clusters).
These clusters could either be lost to the Milky Way bulge or be later captured by the GC, depending on their location in the Galactic bulge and the tidal field.
The more favourable interpretation of the dynamical evolution of the clusters is that these new stars would be captured by GC so as to increase the number of new stars within its tidal radius.

Fig. 10 show that the density of the old stars is comparable to that of the fiducial model. 
New stars are not as localised to the centre of the cluster, partly because a significant fraction of new stars are within sub-clusters formed 
outside the GC.
This model shows a higher efficiency in converting gas into stars, with $ ~78\% $ of the gas particles having been transformed into new stars, more than double the efficiency of M1.
Further simulations of the evolution of this GC have confirmed that the young cluster outside the main GC seen in the final time step, can merge with the GC within 20 Myr if there is no tidal field of the Galactic bulge. 
Such merging can form the outer stellar halo composed of metal-rich stars in the GC.

The star formation rate of this model is very different to the previous model. There is an extremely high star formation rate of nearly $3  {\rm  M}_{\odot} {\rm yr} ^{-1}$ and the majority of the star formation takes place within 4 Myr. Within 6 Myr the new population of stars within the GC has been formed.  This high star formation rate compared to M1 is a direct result of the increase in density of the GMC. The peak between 7 Myr and 10 Myr is caused by the formation of the second star cluster approximately 50 pc away from the GC's nucleus.

\begin{figure}
\psfig{file=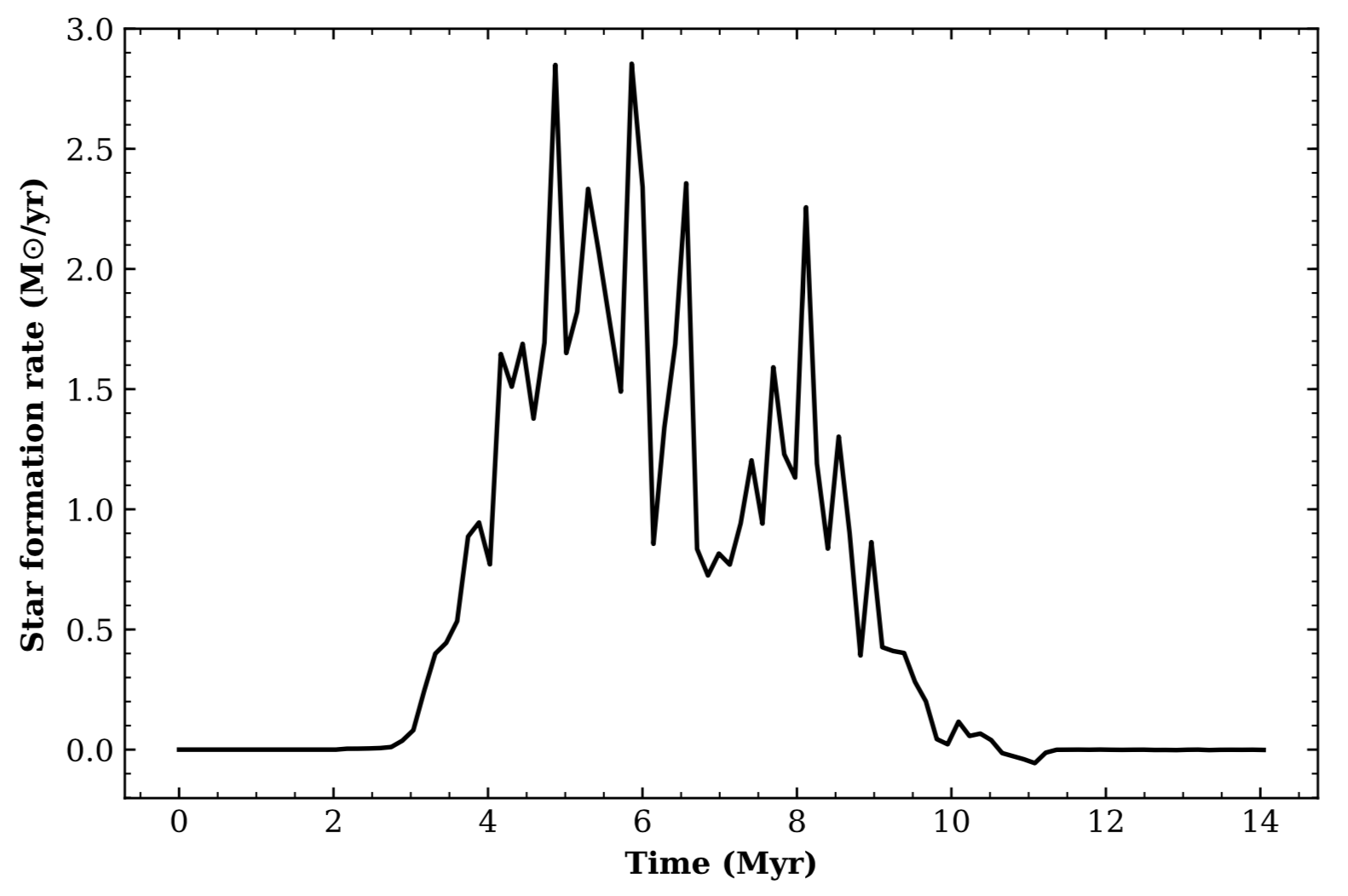,width=8.5cm}
\caption{
The same as Fig. 7 but for M2.
}
\label{Figure.11}
\end{figure}

The cumulative distribution function within 8 pc has also been included in Fig. 12. Compared to M1 there is a more pronounced central concentration of new stars. Most notably, as with M1 this model shows a more notable concave shape within the first three parsecs whereas Ferraro et al. 2009 has a much more convex shape to the distribution. This implies that more of the mass of the cluster is distributed further away from its centre. When we analyse the number of stars within 2 pc as we did for the previous model, we again see this concave shape in our distribution. Within this 2 pc radius, there was a total mass of $9.49\times10^4{\rm M}_{\odot}$ for the old stars and $3.30\times10^4{\rm M}_{\odot}$, almost double the number of new stars compared to the previous model. When the radius is extended to 8 pc we see $1.26\times10^6{\rm M}_{\odot}$ old stars and $0.22\times10^6{\rm M}_{\odot}$. The number of old stars within this radius agrees with observations, but there is a slight deficiency in new stars. An increase in the density of the GMC or the merging of one of the smaller populations generated during the collision could thus increase this number. This leads to the conclusion that the original GC mass of $3.0\times10^6{\rm M}_{\odot}$ is very good at explaining the observed characteristics of the real Ter 5, but there is room to alter the GMC density in order to achieve an optimal number of stars.

\begin{figure}
\psfig{file=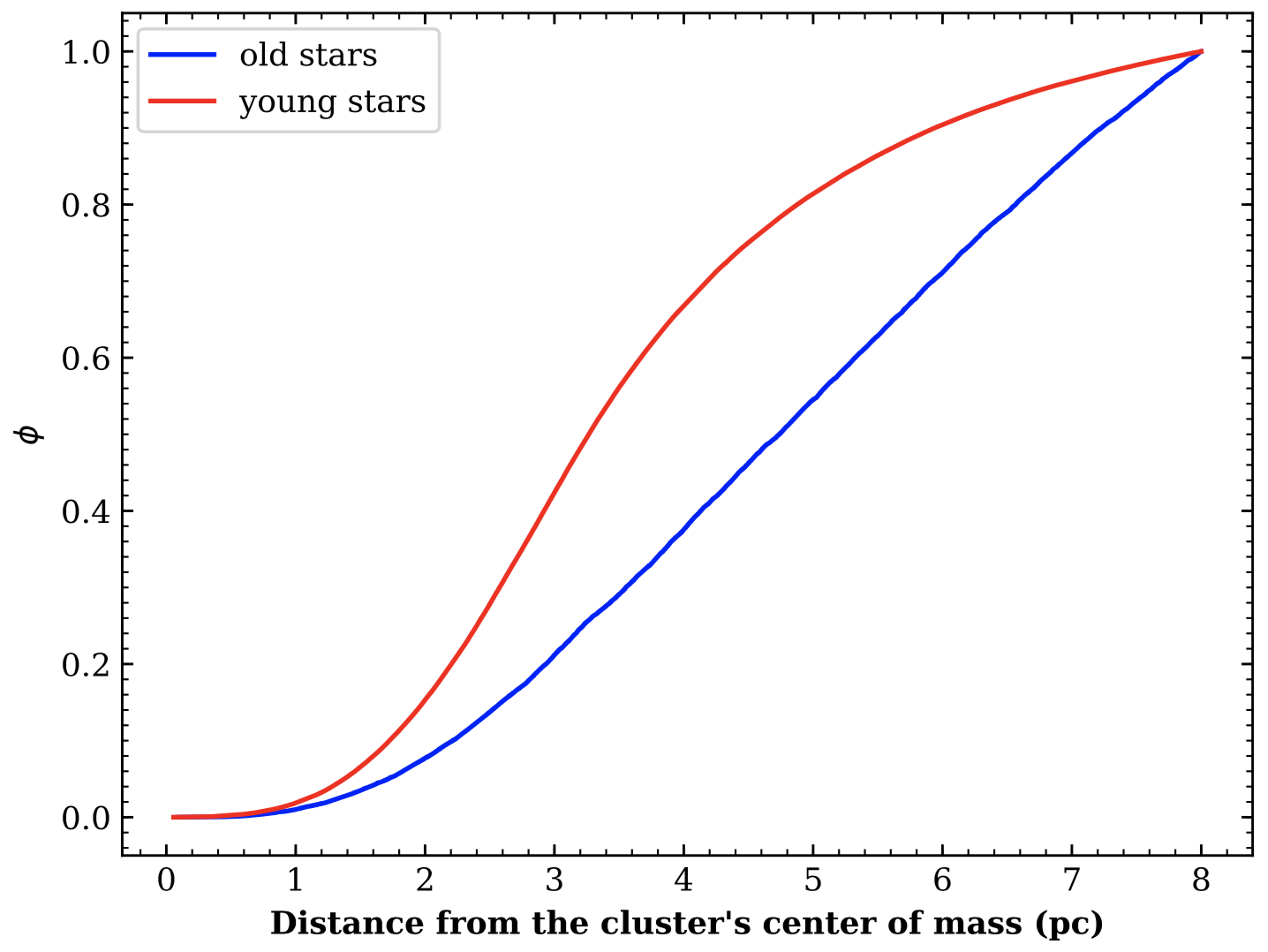,width=8.5cm}
\caption{
The same as Fig. 9 but for M2.
}
\label{Figure.12}
\end{figure}

The velocity dispersion of M2 is shown to be 15.03 Km s$^{-1}$ for the new stars and 13.46 Km s$^{-1}$ for the old stars. Combining both population, there is a total dispersion of 14.87 Km s$^{-1}$. This is much closer to the measured value by Massari (2014) and could point to M2 being a more accurate representation of Ter 5's evolution. The maximum velocity dispersion was found in the very centre of the cluster for both populations.

\subsubsection{Parameter dependence}

We investigated over 100 models in order to identify the effects of each parameter. 
In several models with masses greater than $10^7 {\rm M}_{\odot}$ the GC developed a disk of new stars parallel to its motion.
Low impact parameters and velocities usually resulted in the most realistic models. 
When higher impact parameters were used ($>60$ pc), the new stars started to form an annulus rather than a centrally concentrated disk, thus decreasing the central concentration of stars. 
When the incoming GC velocity was too high, not enough new stars were formed due to the short contact time between the cloud and the GC.

The impact parameter and incoming GC velocity were the main parameters that were changed between simulations. As previously noted, a high impact parameter would result in an annulus rather than a disk and low velocities would create an excessive amount of new stars. We predominantly investigated these lower collisional parameters, thus most models outputted reasonable ratios of new to old stars, depending on the designated tidal radius.

The radius of the GC became central in determining the final mass. A small starting radii resulted in a larger half mass radius owing to the potential energy stored in the initial state of the GC (Table A1, Model 46). Within the first 3 Myrs of the simulation some stars had travelled out to triple their starting radius, thus being stripped from the GC.

The density of the GMC dramatically affected the quantity of new star being formed. 
A high GMC density generated several individual groups of new stars with smaller 
centralised populations and a large percentage being flung out into the GMC. For these models as well as all other models tested, the central collection of new stars always shared the centre of mass of the progenitor GC.
This phenomenon of new stars being launched into the GMC also occurred for 
models with large GC radii and slow velocities in some high resolution simulations. 
In future studies of the Galactic bulge, if small clusters of bulge stars with ages of $\sim 5$ Gyr were observed, 
that would point towards a GC-GMC collision model with a higher GMC density.

The velocity dispersion of several simulations demonstrated that the younger population usually has a higher dispersion than the older population. Low mass clusters tended to have lower dispersion compared to high mass clusters, as an increase in stars results in more variations in motion between and therefore a higher spread in velocities. Higher impact parameters and low velocities will result in higher velocity dispersions.
It should be noted that over the $\approx$4 Gyr after the collision occurs, the initial angular momentum would decrease and the cluster would become increasingly dominated by random motions (e.g., Tiongco et al. 2017). This will lead to an increase in the velocity dispersion of the cluster over time, thus M2 and others like it would be more suitable representations of the cluster as they allow for an increase in the velocity dispersion while still closely matching the observed value. 
The results of the velocity dispersion for multiple simulations have been included in Table 1B.

As shown by the present two-stage numerical simulations, it is possible that only one or a few GMCs among $\sim 1000$ were captured by the GC
to finally be converted into new stars. 
GMCs that were not captured by Ter 5 should also have formed new stars during their evolution in the bulge.
Accordingly, the new stars in Ter 5 should only be a minor fraction of the total number of new stars that formed from GMCs in the bulge. 
It is possible that the younger population of Ter 5 is only $\sim 0.1$\% of all the bulge stars which formed around 5 Gyrs ago. 
This implies that the total mass of $\sim 5$ Gyr old stars in the bulge can be as large as $3 \times 10^8 {\rm M}_{\odot}$ given the total mass of the younger population of Ter 5 is $3 \times 10^5 {\rm M}_{\odot}$. 
This is a significant fraction of the bulge stars and thus supports the scenario that the bulge has younger, metal-rich stellar populations (e.g., Nataf 2016).

\section{Discussion}

\subsection{Advantages and disadvantages of the new scenario}

\subsubsection{Advantages}

The main advantage of our new model is that it can account for the two distinct ages of Ter 5 with a gap of 5 Gyrs, without requiring abundant SNII events in an unusually large stellar system with a very deep gravitational potential.
The previous scenario based on chemical enrichment due to numerous SNII cannot simply explain why such enrichment (and the subsequent star formation from the metal-enriched gas) can occur several Gyrs after the formation of the main metal-poor population.
Our model results in a centrally concentrated distribution of stars with both populations sharing the same centre of mass, and stars are formed in the appropriate ratio of old to new, as per observations.
This self-consistent reproduction of the observed mass distributions of the two distinct stellar populations in Ter 5 can be regarded as a significant
advantage of the new model.
Merging between two clusters with different ages and metallicities can be possible in galaxies (Bekki \& Tsujimoto 2016), however, such merging cannot result in a stronger mass concentration of the younger population.
Many models with a range of parameters generate centrally concentrated distributions of younger, metal-rich stellar populations
in the present simulations, which demonstrates that our results are robust.

\subsubsection{Disadvantages}
A critical factor in our simulations is that we have assumed that a large number (an order of $\sim 1000$) of GMCs were present during the early evolution of our Galaxy. 
Their existence is fundamental to our theory and more galactic simulations should be run in order to support or refute our claims.
Since the Galaxy was more metal-poor and more gas-rich, as chemical evolution models of the Galaxy suggest (e.g., Tsujimoto et al. 2010), the Galactic bulge could have contained a larger fraction of molecular gas.
However, there is no direct observational study that can provide some information on the total mass of GMCs in the central region of the Galaxy
5 Gyr ago.

Additionally, our model relies on a significant proportion of these GMC having a solar-scaled alpha-element abundance in order to generate the metal-rich new population. Nataf (2017) demonstrated using data from Bensby et al. (2017) that a low [$\alpha$/Fe] at metallically [Fe/H] = 0.2 dex is possible for the bulge stars approximately 5 Gyr ago. Therefore, the presence of these metal rich GMCs may not have been that unlikely during the birth of the younger population.

Unlike the self-enrichment model of Ter 5, our model does not directly give evidence for why Ter 5 has such a large millisecond pulsar population;
we can provide a possible explanation for the origin of the pulsar population in the next sub-section.  
Results from the first stage of our simulations were erratic and small changes in parameters could drastically change the number of gravitationally bound particles. 
Additionally, when we did find a gravitationally bound collision, the impact parameter and relative velocity of the collision were often
too large to form new stars. 
This rarity of strongly bound GC-GMC collisions may explain why Ter 5 is such an anomaly in the Galaxy.  
However, if such a bound GC-GMC collision is extremely rare (e.g., owing to the small number of GMCs), then the GC-GMC collision scenario could be ruled out.

\subsection{The origin of the massive progenitor of Terzan 5}

The present study has shown that an initially large mass ($ \ge 3 \times 10^6 {\rm M}_{\odot}$) of Ter 5's progenitor is preferred in the GC-GMC collision scenario, because the total mass of new metal-rich stellar populations formed from gas accretion during GC-GMC collision is large ($>10^5 {\rm M}_{\odot}$) enough to be consistent with observations.
Recent numerical simulations of the Galactic GCs have shown that GCs in the inner region of the Galaxy can lose significant fractions of stars owing to the efficient tidal stripping of the Galaxy (e.g., Rossi et al. 2016 for the latest results).  
Therefore, the initially adopted large mass of Ter 5 (i.e., the initial mass of metal-poor population) is not a problem, given the present-day mass of $2 \times 10^6 {\rm M}_{\odot}$ for Ter 5.

If the original mass of the main metal-poor population in Ter 5
(i.e., Ter 5's progenitor stellar system) is as large as $10^7 {\rm M}_{\odot}$, then it can quickly sink into the Galactic centre due to dynamical friction.
The time scale of such dynamical friction ($t_{\rm fric}$) for a stellar system with a mass of $M_{\rm gc}$ 
is as follows:
\begin{equation}
t_{\rm fric}= 2.64
{ ( \frac{ \ln \Lambda} {10} )  }^{-1}
{ ( \frac{ r_{\rm i} } {2 {\rm kpc} } )  }^{2}
{ ( \frac{ v_{\rm c} } {250 {\rm km/s} } )  }
{ ( \frac{ M_{\rm gc} } {10^7  {\rm M}_{\odot} }  )  }^{-1}
{\rm Gyr},
\end{equation}
where $r_{\rm i}$ is the initial distance of the GC from the galaxy centre,
and $v_{\rm c}$ is the maximum circular velocity of the Galaxy,
and $\Lambda$ is the coulomb logarithm.
This implies that Ter 5's progenitor should have spiralled into the Galactic centre.

Since Ter 5 still exists in the bulge area of the Galaxy, a way to avoid such a very short time scale of dynamical friction
is to assume that Ter 5's progenitor was initially located outside of the bulge (e.g., $R>2$ kpc). For example, if the
Ter 5's progenitor was initially located at $R \sim 6$ kpc, then
$t_{\rm fric}$ can be $\sim  23$ Gyr, which is longer than Hubble time.
Therefore,  the progenitor stellar system has not yet spiralled into the Galactic centre.
This suggests that the progenitor system was formed outside the bulge and has recently migrated into the bulge region. 
Accordingly, if the progenitor's mass is really as massive as $10^7 {\rm M}_{\odot}$, then the progenitor needs to be born well outside the bulge.
Such a massive progenitor of Ter 5 formed well outside the bulge can be either stellar nucleus of a defunct building block of the Galaxy or a massive clump (or fragment) of the early Galactic disk. 
Further studies of our model will discuss which of the two candidates for Ter 5's progenitor are more realistic.

\subsection{The origin of the metal-poor minor population}

Ter 5 is observed to have an additional ``third''  stellar population
with a metallicity of [Fe/H]$\sim -0.79$ (Origlia et al.2013). This collection of stars is much smaller and much more metal deficient than the main metal-poor population which was previously discussed.
More simulations could be run of the progenitor stellar system of Ter 5 in order to shed light on the third metal-poor component. 
However, we here can suggest that Ter 5 was originally a member of a dwarf galaxy which was destroyed by our own Galaxy. What we previously labelled the progenitor GC could potentially be the product of two systems which merged together each in their own dwarf galaxy.
This remnant of GC-GMC merging could then be accreted by the Milky Way and go on to eventually collide with a GMC.
Because of the relative low velocities of GCs in dwarf galaxies, a merging event could occur within them (Bekki \& Tsujimoto 2016).
Thus, the third population could have originally been a low-mass cluster that merged with Ter 5 when they were both hosted by a defunct dwarf galaxy.

\subsection{The origin of the millisecond pulsar population}

One piece of key supporting evidence of the self-enrichment 
theory is Ter 5's abundance of millisecond pulsars. 
Our model can explain the observed number of millisecond pulsars as follows.
GCs with younger ages are observed to have a larger fraction
of  binary star systems (e.g., Ji \& Bregman 2015). 
Our new simulations have shown that the young metal-rich population can be centrally concentrated within the existing old and metal-poor
population. 
These observational and theoretical results therefore suggest
that the central region of Ter 5 can have a larger fraction of binary stars,
which can result in the formation of binary neutron stars (i.e., progenitor
of millisecond pulsars).
It is beyond the scope of this paper to discuss quantitatively how
many millisecond pulsars can be formed in the young metal-rich stellar
population of Ter 5.
In future studies we will investigate whether the observed number of millisecond pulsars in Ter 5 can be explained by our new scenario.

\section{Conclusions}

We have proposed a new formation scenario of Ter 5 (``GC-GMC collision'') in which the metal-rich ([Fe/H]$\sim +0.2$), younger ($\sim 5$ Gyr old) stellar population was formed from a collision between a metal-rich GMC in the Galactic bulge and the metal-poor massive progenitor of Ter 5 about 5 Gyr ago.
We have used this ``GC-GMC collision'', even though the original stellar system of Ter 5 was much larger than the typical mass
($2 \times 10^5 {\rm M}_{\odot}$) of the Galactic GCs.
In the new scenario, new metal-rich stars are assumed to be formed from gas accreted onto the
original massive stellar system of Ter 5 during the GMC-GC collision.

The model draws on the idea of Ter 5 being a metal-poor 
massive stellar system in the bulge, where the mass of the new generation of stars is
directly sourced from the same gas that the younger bulge stars are  generated from. 
Using dynamical and hydrodynamical simulations,
we have first investigated whether such a GC-GMC collision
was possible in the bulge region of the younger Galaxy about 5 Gyr ago.
We then have investigated the details of the GC-GMC collision for a wide
range of model parameters in order to confirm that metal-rich stars
can be formed from gas of a colliding GMC.
We have investigated numerous models with
a wide spread in GC masses and radii and a range of GMC 
densities.
The principal results of this two-fold investigation are summarised as follows.

(1) Although Ter 5 can collide with numerous GMCs in the Galactic bulge
($R<3$ kpc from the Galactic centre) over a time-scale of a 
few Gyrs, almost all of the collisions are high-speed ones with the relative
velocities being higher than the escape velocity of Ter 5. Therefore,
no/little gas accretion is possible in these high-speed collisions.
However,  GC-GMC collisions in which a GMC can be gravitationally
captured by Ter 5 (`` bound collisions'')
are possible for a range of model parameters.

(2)  Such bound GC-GMC collisions are found in the models in which 
Ter 5 is massive and has a co-planer orbit with respect to the Galactic disk plane.
The Galactic bulge can be a key factor that determines whether Ter 5 can 
merge with a GMC in the bulge. 
If the mass of the bulge (more than 5 Gyr ago) is small, then Ter 5 can more frequently merge with a metal-rich GMC within a few Gyrs. 
These results demonstrate that a bound collision between a metal-rich GMC and Ter 5 is possible in the Galactic bulge about 5 Gyr ago.

(3) The GC-GMC collision scenario cannot simply
explain why 5 Gyr ago is a special epoch for Ter 5 
(i.e., why a bound GC-GMC collision was possible for Ter 5 about 5 Gyr ago),
because there is no particular (preferred) epoch for GC-GMC bound collisions
in the scenario.
A possible explanation for the GC-GMC
collision about 5 Gyr ago is that
a stellar bar was formed in the Galaxy
about $5$ Gyr ago and subsequently induced rapid gas
inflow to the Galactic bulge, which ended up with a high molecular gas density
and thus a higher probability of GC-GMC bound collisions.
If the molecular gas density of the bulge region was indeed significantly higher
about 5 Gyr ago, then the present-day Galactic  bulge should have 
stars with ages of $\sim 5$  Gyr.
Thus, the younger population of Ter 5  has some implications 
on the long-term star formation history of the bulge (e.g., Nataf 2016).

(4) A collision between Ter 5 and GMC can create a new, centrally concentrated generation of stars from gas in the GMC. 
However, both the impact parameter and the relative velocity between the GMC and Ter 5 must be low so that gas of the GMC can be accreted onto Ter 5.
The initial mass of Ter 5 should be at least as large as $[3-10] \times 10^6 {\rm M}_{\odot}$ for the formation of young metal-rich stellar populations with the total mass of $\sim 10^5 {\rm M}_{\odot}$ in the central region of Ter 5.

(5) The simulated cumulative mass distributions of existing old stars and
new ones can be consistent with corresponding observations in some models
of the present study. The more centrally concentrated younger population
of Ter 5 can be naturally reproduced by the present GC-GMC collision
scenario, because gaseous dissipation during the GC-GMC collision can
lead to the formation of a compact stellar system surrounded by the initially
existing stellar system of the GC. Global rotation of new metal-rich stars
in the simulated stellar systems are not clear in the present study.
More observational data about the internal rotations of Ter 5 will benefit our understanding about the internal kinematics of the two populations.

(6) We have demonstrated the two best models for the formation
of Ter 5: one is that the original metal-poor
progenitor stellar system  of Ter 5 
was originally as massive as $10^7 {\rm M}_{\odot}$
(a large difference between the present-day and original masses of Ter 5).
The other is that it started off at roughly it's current size, 
underwent minimal stripping and went on to form what it is today. 
We have suggested that for the former case,
Ter 5 needed to have underwent tidal stripping by the Galactic bulge
and have lost most of its mass within 5 Gyrs.

(7)  We have provided physical reasons for the existence of the large number of millisecond pulsars and the 
metal-poor ([Fe/H]$\sim -1.2$) minor (``third'') population in this GC-GMC collision scenario.

Further simulations of the early evolution of the Milky Way Galaxy
will help justify the existence of many GMCs in the central region of the Galaxy.
This model brings us one step closer to a unified understanding of Ter 5 and the events it underwent during its evolution.

\section*{Acknowledgements}

We are grateful for the referee's thoughtful and constructive comments. MM thanks ICRAR for the opportunity to have participated in their 2018 studentship.

\appendix

\section{Results of less successful models}

\begin{table*}
\centering
\begin{minipage}{170mm}
\caption{Parameters values and results for a sample of the test models that were investigated. 
Most models are unsuccessful in reproducing the observed properties of Ter 5 and thus are not described
in the main text.}
\begin{tabular}{cccccccccc}\hline
Model ID & 
{ $M_{\rm gc}$   ($10^6 {\rm M}_{\odot}$)   } & 
{ $R_{\rm gc}$ (pc)  } &
{ $b$ (pc)  } &
{ $V_{\rm x}$ (km/s)   } & 
{ $M_{\rm gmc}$   ($10^6 {\rm M}_{\odot}$)   }  &
{ $R_{\rm gmc}$ (kpc)  }  & 
{ $M_{\rm old}$   ($10^6 {\rm M}_{\odot}$)   }  & 
{ $M_{\rm new}$   ($10^6 {\rm M}_{\odot}$)   }  &  \\\hline
1  & 10 & 50 & 80 & 30  & 1   & 0.1  & 3.2  & 0.04 \\
2  & 3  & 5  & 94 & 9.7 & 1   & 0.1  & 0.34 & 0.39 \\
3  & 3  & 5  & 10 & 9.7 & 1   & 0.1  & 0.65 & 0.4  \\
4  & 3  & 5  & 20 & 9.7 & 1   & 0.1  & 0.84 & 0.35 \\
5  & 3  & 5  & 20 & 20  & 1   & 0.1  & 0.14 & 0.10  \\
6  & 3  & 5  & 20 & 5   & 1   & 0.1  & 0.62 & 0.36 \\
7  & 3  & 5  & 10 & 5   & 1   & 0.1  & 0.55 & 0.44 \\
8  & 3  & 5  & 20 & 5   & 1   & 0.1  & 0.62 & 0.36 \\
9  & 3  & 5  & 5  & 10  & 1   & 0.1  & 0.55 & 0.41 \\
10 & 8  & 12 & 5  & 10  & 1   & 0.1  & 1.95 & 0.67 \\
11 & 10 & 15 & 5  & 10  & 1   & 0.1  & 2.39 & 0.70  \\
12 & 12 & 18 & 7  & 10  & 1   & 0.1  & 2.87 & 0.75 \\
13 & 12 & 20 & 10 & 12  & 1   & 0.1  & 3.45 & 0.70  \\
14 & 15 & 25 & 10 & 12  & 1   & 0.1  & 0.41 & 0.73 \\
15 & 15 & 25 & 15 & 12  & 1   & 0.1  & 4.27 & 0.72 \\
16 & 15 & 25 & 15 & 18  & 1   & 0.1  & 3.98 & 0.56 \\
17 & 15 & 30 & 18 & 10  & 1   & 0.1  & 4.85 & 0.75 \\
18 & 15 & 30 & 15 & 12  & 1   & 0.1  & 4.52 & 0.73 \\
19 & 15 & 25 & 20 & 5   & 1   & 0.1  & 3.88 & 7.95 \\
20 & 15 & 35 & 25 & 5   & 1   & 0.1  & 4.49 & 0.83 \\
21 & 12 & 35 & 25 & 5   & 1   & 0.1  & 3.83 & 0.82 \\
22 & 10 & 35 & 25 & 5   & 1   & 0.1  & 3.72 & 0.76 \\
23 & 10 & 35 & 35 & 5   & 1   & 0.1  & 3.81 & 0.65 \\
24 & 10 & 35 & 35 & 8   & 1   & 0.1  & 3.97 & 0.52 \\
25 & 8  & 35 & 40 & 5   & 1   & 0.1  & 3.43 & 0.61 \\
26 & 8  & 35 & 20 & 10  & 1   & 0.1  & 3.37 & 0.62 \\
27 & 10 & 35 & 20 & 10  & 1   & 0.1  & 3.93 & 0.65 \\
28 & 10 & 30 & 20 & 10  & 1   & 0.1  & 3.85 & 0.67 \\
29 & 10 & 25 & 15 & 10  & 1   & 0.1  & 3.37 & 0.73 \\
30 & 10 & 25 & 15 & 5   & 1   & 0.1  & 2.81 & 0.79 \\
31 & 10 & 35 & 35 & 5   & 1   & 0.1  & 3.81 & 0.65 \\
32 & 5  & 18 & 35 & 5   & 3   & 0.1  & 0.99 & 2.00    \\
33 & 5  & 10 & 30 & 5   & 3   & 0.1  & 0.49 & 2.03 \\
34 & 5  & 30 & 10 & 5   & 3   & 0.1  & 1.05 & 2.26 \\
35 & 10 & 30 & 10 & 5   & 3   & 0.1  & 1.62 & 2.33 \\
36 & 10 & 30 & 50 & 8   & 3   & 0.1  & 3.08 & 1.79 \\
37 & 3  & 5  & 50 & 8   & 3   & 0.1  & 0.51 & 1.62 \\
38 & 3  & 5  & 70 & 8   & 3   & 0.1  & 0.26 & 1.71 \\
39 & 5  & 8  & 40 & 3   & 3   & 0.1  & 0.53 & 0.94 \\
40 & 5  & 8  & 40 & 3   & 3   & 0.1  & 0.21 & 1.92 \\
41 & 3  & 5  & 40 & 3   & 3   & 0.1  & 0.13 & 2.00    \\
42 & 3  & 5  & 40 & 5   & 3   & 0.1  & 0.23 & 1.88 \\
43 & 10 & 20 & 20 & 8   & 1   & 0.06 & 1.77 & 0.54 \\
44 & 10 & 20 & 20 & 40  & 1   & 0.06 & 2.19 & 0.33 \\
45 & 10 & 20 & 20 & 15  & 1   & 0.06 & 2.28 & 0.43 \\
46 & 3  & 1  & 20 & 5   & 1   & 0.06 & 0.00    & 0.00    \\
47 & 3  & 20 & 30 & 10  & 1   & 0.06 & 1.36 & 0.28 \\
48 & 3  & 20 & 50 & 10  & 1   & 0.06 & 1.48 & 0.04 \\
49 & 3  & 20 & 90 & 8   & 1   & 0.06 & 1.34 & 0.00    \\
50 & 3  & 20 & 30 & 12  & 1   & 0.06 & 1.37 & 0.25 \\
51 & 3  & 20 & 30 & 12  & 1.5 & 0.06 & 1.00    & 0.32 \\
52 & 3  & 20 & 30 & 10  & 1.5 & 0.06 & 0.88 & 0.39 \\
53 & 2  & 20 & 30 & 10  & 1   & 0.06 & 0.98 & 0.17 \\
54 & 3  & 20 & 55 & 13  & 2   & 0.06 & 0.90  & 0.01 \\
55 & 3  & 20 & 74 & 13  & 2   & 0.06 & 0.65 & 0.15 \\
56  & 2  & 20 & 35 & 5  & 1   & 0.06 & 1.00    & 0.05 \\
57  & 12 & 18 & 7  & 10 & 1   & 0.1  & 2.86 & 0.75 \\
58  & 6  & 10 & 5  & 10 & 1   & 0.1  & 1.30  & 0.61 \\
59 & 12 & 18 & 5  & 10 & 1   & 0.1  & 3.22 & 0.72 \\
60  & 12 & 15 & 10 & 10 & 1   & 0.1  & 2.86 & 0.71 \\
61  & 10 & 25 & 10 & 5  & 1   & 0.1  & 2.67 & 0.75 \\
62  & 5  & 20 & 30 & 5  & 3   & 0.1  & 1.01 & 2.01 \\
63 & 5  & 30 & 20 & 5  & 3   & 0.1  & 1.00    & 2.17 \\
64 & 3  & 20 & 30 & 5  & 1   & 0.06 & 0.95 & 0.21 \\
65 & 2  & 20 & 20 & 5  & 1.5 & 0.06 & 0.67 & 0.11
\end{tabular}
\end{minipage}
\end{table*}

We have shown the results of the two representative models, M1 and M2 in the main text as they both show typical behaviours of Ter 5 formation with metal-rich young stars from a GC-GMC collision.
It should be noted here that we have run many
low-resolution ($N<2\times10^5$) models in order to find the best model(s) that
can reproduce the observed properties of Ter 5.
It is found that (i) there are many {\it unsuccessful}
models that failed to explain the physical properties of Ter 5
(e.g., the total mass of metal-rich stars) and (ii) a number of models
can reproduce the observed cumulative mass distributions 
of old and younger stars in Ter 5 to some extent.
Below, we briefly summarise the results of these less important models.

(1) Initial low-mass GCs ($ \sim 3 \times 10^6 {\rm M}={\odot}$)
such as M2-9, M37-42 and M46 result in smaller total masses ($<10^6 {\rm M}_{\odot}$)of old stars after GC-GMC collisions. This is a result of a significant fraction
of old stars of the GCs being lost during tidal interaction between the GCs and the GMCs.
However, these models show significant final masses of SG stars, which means 
that the mass-ratio of new (SG)  stars to old stars in each  
of these models  is not so consistent with observations
(i.e., the simulated mass-ratio is a bit too large).

(2) The relative velocity between a GMC and a GC
does not make a major difference in the final mass of old stars within the tidal radius 
of the GC. This is because GC have larger masses and higher densities
than GMC in almost all models (so that GCs are less influenced by the gravitational interaction
with GMCs).
A large relative velocity does however decrease the total mass
of new stars formed from a GMC (e.g. M49). This is mainly because only a smaller fraction of gas
in the GMC can be accreted onto the colliding GC.

(3) The impact parameter for a GC-GMC collision
is less influential on the total mass of new stars formed
from the GMC within the GC,
but it has a greater impact on the final mass distribution of old and new stars.
A larger impact parameter can decrease the centrality of the new stars
as previously mentioned in \S 3,
because only a smaller fraction of gas in a GMC can be captured by
the colliding GC.

(4) An increase in GMC density whereby the mass is increased from $10^6$ to $3\times10^6$ but radius remains constant results in a decrease in old stars due to dynamical friction (see M32 to M42). When the radius is decreased in models M43 to M56, the mass of the old stars remains relatively constant, but we also see an increase in new star mass. This is beneficial for lower mass clusters as discussed previously.

\section{Velocity dispersion of both populations of the simulated GC}
A summary of the rotation of selected models has been included in Table B1. Here M1 and M2 correspond to the same M1 and M2 used in \S 3. The other models included were at a higher resolution than those included in Table A1 as more particles were needed to obtain an accurate reading of the velocity dispersion. Thus the rest of the models listed are not related to models in other tables. As mentioned in the main text, many low mass models required higher density GMCs to accrete an optimal amount of gas. As this directly influences star formation and thus velocity dispersion of the stars, these high density GMC models are indicated by a $*$ in the table.

\begin{table*}
\centering
\begin{minipage}{165mm}
\caption{Velocity dispersions ($\sigma$) for various models.  * denotes models which used a high density GMC}
\label{my-label}
\begin{tabular}{cccccccc}\hline
{Model ID} & 
{ $M_{\rm gc}$   ($10^6 {\rm M}_{\odot}$)   } & 
{ $R_{\rm gc}$ (pc)  } &
{ $b$ (pc)  } &
{ $V_{\rm x}$ (km s$^{-1}$)   } &  
{ $\sigma_{\rm new}$ (km s$^{-1}$)   }  & 
{ $\sigma_{\rm old}$ (km s$^{-1}$)   } & 
{ $\sigma_{\rm total}$ (km/ s$^{-1}$)   }  \\\hline
1        & 10   & 20     & 40          & 20      & 19.04     & 19.36     & 19.07       \\
2*       & 3    & 20     & 20          & 10      & 15.03     & 13.48     & 14.87       \\
3        & 3    & 20     & 20          & 10      & 16.81     & 14.35     & 15.98       \\
4*       & 3    & 20     & 30          & 10      & 18.73     & 13.99     & 15.27       \\
5*       & 2    & 20     & 35          & 5       & 12.43     & 11.57     & 11.74       \\
6        & 10   & 35     & 35          & 5       & 25.96     & 20.72     & 22.67       \\
7        & 10   & 20     & 35          & 10      & 23.31     & 19.69     & 21.25       \\
8        & 10   & 20     & 20          & 20      & 18.46     & 19.38     & 18.75       \\
9        & 10   & 20     & 35          & 20      & 21.82     & 19.27     & 20.86       \\
10*      & 3    & 20     & 30          & 10      & 21.70     & 14.10     & 16.30       \\
11*      & 3    & 20     & 20          & 12      & 25.54     & 15.08     & 18.88       \\
12       & 3    & 20     & 20          & 12      & 14.86     & 14.25     & 14.48      
\end{tabular}
\end{minipage}
\end{table*}

\end{document}